\documentclass[10pt]{article}

\usepackage{geometry}
\usepackage{ amsmath, amsfonts, amsthm,  amssymb, url, mathrsfs}
\usepackage{relsize}
\usepackage{color, tikz, pgfplots, todonotes, amssymb, mathrsfs}
\usepackage{framed}
\usepackage[colorlinks=true, pdfstartview=FitV, linkcolor=darkblue, citecolor=darkblue, urlcolor=darkblue]{hyperref}%
\usepackage[nobottomtitles*]{titlesec}

\def\bea#1\eea{\begin{align} #1 \end{align}}
\def \bs{\boldsymbol}
\def\bd{ \begin{definition}}
\def\ed{\end{definition}  }
\def\bt{  \begin{theorem}\rm }
\def\et{\end{theorem}  }
\def \d{\mathrm d}

\def\nn{\nonumber}
\def\bp{ \begin{proposition}}
\def\ep{\end{proposition} }
\def\s{\sigma}
\def \Lie{ \LL}

\definecolor{shadecolor}{rgb}{0.98, 0.98, 0.9}
\definecolor{darkgreen}{rgb}{0.2, 0.5,  0}
\definecolor{darkblue}{rgb}{0.1,0.1,0.45}
\definecolor{red}{rgb}{0.9,0,0}
\definecolor{blue}{rgb}{0.,0.,0.75}

\def \C{\mathbb C}
\def \L{\Lambda}
\def \1{\mathbf 1}

\newtheorem{theorem}{Theorem}[section]
\newtheorem{lemma}[theorem]{Lemma}
\newtheorem{definition-theorem}[theorem]{Definition-Theorem}
\newtheorem{definition-proposition}[theorem]{Definition-Proposition}

\newtheorem{proposition}[theorem]{Proposition}
\newtheorem{corollary}[theorem]{Corollary}
\newtheorem{example}{Example}[section]
\newtheorem{examples}{Example}[subsection]

\newtheorem{remark}{Remark}[section]

\newtheorem{definition}{Definition}[section]

\numberwithin{equation}{section} % requires package amsthm

\def\m{\mathop}
\def \l{\lambda}
\def \pa{\partial}
\def\ra{{\rightarrow}}

\def\wt{\widetilde}
\def\Tr{\mathrm {Tr}}
\def\tr{\mathrm {tr}}
\def\det{\mathrm {det}}

\def\Ad{\mathrm {Ad}}
\def\ad{\mathrm {ad}}

\def\Span{\mathrm {Span}}
\def\Diag{\mathrm {Diag}}
\def\diag{\mathrm {diag}}
\def\ss{\subset}

\def\res{\mathop{\mathrm {res}}\limits}
\def\ds{\displaystyle}
\def\un{\underline}
\def\&{&{\hskip -20pt}}

\renewcommand{\le}{\left}
\newcommand{\ri}{\right}
\def \wh{\widehat}
\def \ad{\mathrm{ad}}

\def\be#1\ee{\begin{align} #1 \end{align}}

\def\bea{\begin{eqnarray}}
\def\eea{\end{eqnarray}}

\def\bex{\begin{example}\small \rm}
\def\eex{\end{example}}

\def\bexs{\begin{examples}\small \rm}
\def\eexs{\end{examples}}

\def\br{\begin{remark}\small \rm}
\def\er{\end{remark}}

\def\CC{{\mathcal C}}
\def\DD {{\mathcal D}}

\def\II {{\mathcal I}}

\def\LL{{\mathcal L}}

\def\OO{{\mathcal O}}
\def\PP{{\mathcal P}}

\def\TT {{\mathcal T}}

\def\XX{{\mathcal X}}
\def\YY{{\mathcal Y}}

\def\Ib{\mathbf{I}}

\def\Tb{\mathbf{T}}

\def\Wb{\mathbf{W}}
\def\Xb{\mathbf{X}}

\def\cb{\mathbf{c}}
\def\db{\mathbf{d}}

\def\tb{\mathbf{t}}

\def\Cbb{\mathbb{C}}

\def\Nbb{\mathbb{N}}
\def\Pbb{\mathbb{P}}

\def\Zbb{\mathbb{Z}}

 \def\grh{\mathfrak{h}}
\def\grI{\mathfrak{I}}

 \def\grl{\mathfrak{l}}

 \def\grs{\mathfrak{s}}

\def\grGl{\mathfrak{Gl}} \def\grgl{\mathfrak{gl}}

\renewcommand{\theequation}{\arabic{section}.\arabic{equation}}

\makeatletter
\@addtoreset{equation}{section}
\makeatother

\begin{document}
\baselineskip 16pt

\medskip
\begin{center}
\begin{Large}\fontfamily{cmss}
\fontsize{17pt}{27pt}
\selectfont
	\textbf{Hamiltonian structure of  rational isomonodromic deformation systems}
	\end{Large}
	
\bigskip \bigskip
\begin{large} 
M. Bertola$^{1,2,3}$\footnote[1]{e-mail:marco.bertola@concordia.ca},
J. Harnad$^{1, 2}$\footnote[2]{e-mail:harnad@crm.umontreal.ca} 
and
J. Hurtubise$^{1,4}$\footnote[3]{e-mail:jacques.hurtubise@mcgill.ca} 
 \end{large}
 \\
\bigskip
\begin{small}
$^{1}${\it Centre de recherches math\'ematiques, Universit\'e de Montr\'eal, \\C.~P.~6128, succ. centre ville, Montr\'eal, QC H3C 3J7  Canada}\\
$^{2}${\it Department of Mathematics and Statistics, Concordia University \\ 1455 de Maisonneuve Blvd.~W.~Montreal, QC H3G 1M8  Canada}\\
$^{3}${\it SISSA, International School for Advanced Studies, via Bonomea 265, Trieste, Italy }\\
$^{4}${\it Department of Mathematics and Statistics, McGill University \\ 805 Sherbrooke St.~W.~Montreal, QC  H3A 0B9 Canada\\ }
\end{small}
 \end{center}
\medskip
\begin{abstract}
\smaller{
The Hamiltonian approach to isomonodromic deformation systems is extended to include generic rational covariant derivative
operators  on the Riemann sphere with irregular singularities of arbitrary Poincar\'e rank. The space of rational connections with given pole degrees carries a natural Poisson structure corresponding to the standard classical rational R-matrix structure on the dual space $L^*\grgl(r)$ of the loop algebra $L\grgl(r)$. Nonautonomous isomonodromic counterparts of the isospectral systems generated by spectral invariants are obtained by identifying the deformation parameters as Casimir elements on the phase space. These are shown to coincide with the higher {\it Birkhoff invariants} determining the local asymptotics near to irregular singular points,
together with the pole loci.  Pairs consisting of Birkhoff invariants, together with the corresponding {\it dual } spectral invariant Hamiltonians, appear as ``mirror images'' matching, at each pole, the  negative power coefficients in the principal part  of the Laurent expansion of the fundamental meromorphic differential  on the associated spectral curve  with the corresponding positive power terms in the analytic part.  Infinitesimal isomonodromic  deformations are shown to be generated by the sum of the Hamiltonian vector field  and an {\em explicit derivative} vector field that is transversal to the symplectic foliation. The Casimir elements serve as coordinates complementing those along the symplectic leaves, defining a local symplectomorphism between them.  
The  explicit derivative vector fields preserve the Poisson structure and  define a flat {\it transversal connection}, spanning an integrable distribution 
whose leaves may be identified as the orbits of a free abelian local group action.  The projection  of the infinitesimal isomonodromic deformation vector fields  to the quotient manifold  under this action gives the commuting Hamiltonian vector fields corresponding to the spectral invariants dual to the Birkhoff 
invariants and the pole loci.}
\end{abstract}
     \break
     %%%%%%%%%%%%%%%%%%  Table of contents %%%%%%%%%%
     %\tableofcontents
\bigskip

%%%%%%%%%%%%  Section 1. Introduction. Rational Lax matrices and isomonodromic deformations %%%%%%

\section{Introduction and results}
\label{rational_Lax_systems}

%%%%%%%%%%%%%%%  Subsection 1.1  Isomonodromic systems and Hamiltonian structures  %%%%%%%%%%%%%%
\subsection{Isomonodromic systems and Hamiltonian structures}
\label{isomon_ham}

  The study of isomonodromic deformations of linear differential equations with a finite number of isolated singular
   points dates back to the earliest works of Painlev\'e  \cite{Pa1, Pa2, Pa3}, Fuchs \cite{Fu1, Fu2},  Garnier \cite{Gar1}-\cite{Gar3}, 
   Schlesinger  \cite{Sch} and others \cite{Pic, Gam}. A significant extension of what is meant by the
    {\it  generalized monodromy data}  for systems with  irregular isolated singularities was made by Birkhoff \cite{Birk}, who  also
    included  the Stokes and connection matrices.   A revival of interest  in isomonodromic deformations was
stimulated by the work of  Flaschka and Newell \cite{FN1, FN2}  and Jimbo, Miwa and Ueno \cite{JMU, JM}, inspired by 
developments in the theory of  completely integrable systems \cite{H5}. 
        
      Many subsequent studies were devoted to the six Painlev\'e transcendents \cite{Ok1, JM, Ok2, IN, JoKr, CM}, 
   the Schlesinger system \cite{Sch, JMU, H2, Hit}, governing Fuchsian first order matrix systems of arbitrary rank,  
   and generalizations \cite{HTW, H2, Wood1,  Bo2, Bo3, Wood2} in which one or more irregular singular points are present,
    including polynomial systems \cite{FN1, Ug, MaMo}, the Garnier system \cite{Gar1, Gar3}, \cite{HB}, Sec. 9.3,  
    and many other particular cases. The general analysis of \cite{JMU, JM} provides 
   a uniform approach to rational first order systems and has led to a burgeoning literature on their symmetries \cite{Ok1, Ok2, Obl},  
   asymptotic properties \cite{IN, JoKr, IFK, Jo, Gu1}, special classes of solutions, and a variety of applications \cite{HI, CM}. 
  (See \cite{IKSY} for an elementary introduction, \cite{HB}, Chapts.~9-11 and \cite{HTW, BEH1, BEH2, BEH3, HI} 
  for applications to the spectral statistics of random matrices and \cite{LT, ILT, Gu2, GL, ILP} 
  for compendia of various classes of  known solutions.)
   
   One feature that was recognized since the very earliest studies \cite{Malm, Ok1, JMU, JM, Ok2, H2, Bo1, Bo2, IP}
    is that many of these systems could be interpreted as having an underlying Hamiltonian structure of non-autonomous type.  Closely linked to this property is the 
   notion of {\it isomonodromic $\tau$-functions} \cite{JMU, JM}, which allow an alternative representation of these systems in a form that 
 resembles the bilinear Hirota equations \cite{Sa, SW, Hir, H3} characterizing integrable hierarchies of autonomous systems,   both finite and infinite dimensional.   A remarkable feature was that for many classes of isomonodromic systems,  including the Painlev\'e transcendents, the Schlesinger systems and some generalizations of the latter, 
 the  partial derivatives of  the $\tau$-function with respect to the  deformation parameters could be identified with 
 the nonautonomous Hamiltonian functions generating the deformation dynamics, evaluated on the solution manifold.
   
  In this work, previous results on the nonautonomous Hamiltonian structure of such systems
   are extended within a uniform framework that includes all the rational isomonodromic deformation 
   systems  introduced in \cite{JMU, JM}.   This is based on the well-known {\it classical $R$-matrix} structure  \cite{RS1, Sem, RS2} 
   of  rational, linear type, on (the dual space of)  loop algebras. It is the same Poisson structure that plays a 
   fundamental r\^ole in the analysis and solution of  autonomous finite dimensional completely integrable 
   Hamiltonian systems of isospectral type in terms of abelian functions \cite{AvM, AHH1, AHH2, H1}.  
 
  In \cite{HTW, H2} it was noticed that certain classes of isomonodromic deformation equations could be derived 
    using the same phase space and Hamiltonian  structure, by simply treating the systems as 
    nonautonomous deformations of corresponding isospectral ones. Rather than viewing the deformation parameters as independent  time parameters, representing  the evolution under a complete set of commuting flows, they are identified with certain  specific functions on the  phase space which, as it turns out,  are always Casimir elements of the  underlying classical $R$-matrix structure.   
    For the special case of the Schlesinger systems,  this is exactly the converse of the procedure  
   applied by Garnier \cite{Gar2} (which he called the ``Painlev\'e\ simplification'')
    in order to convert  these into autonomous systems that are completely integrable.\footnote{We now recognize these as the classical limit of the general rank-$r$ Gaudin systems \cite{Gau},  which were studied much later as models for quantum integrable spin chains. }
    
    The main message of the present work is that if we do the reverse,  namely {\em deautonomize} a certain subset of the integrable 
    isospectral Hamiltonian systems with respect to the rational classical R-matrix structure,  by identifying the deformation parameters 
    with Casimir elements on the phase space,  we obtain deformed Hamiltonian systems 
    that  agree exactly with the isomonodromic deformation dynamics  introduced in \cite{JMU, JM}.
    Strictly speaking, these are not really Hamiltonian systems since,  added to the Hamiltonian vector fields, there is a ``transversal component'' 
    which amounts  to an ``explicit'' derivative of the corresponding rational Lax matrix $L(z)$ with respect to the Casimir parameters.
    This renders the system no longer isospectral, as a purely Hamiltonian flow would be, but rather of ``zero-curvature'' type,
    on the extended space consisting of the deformation parameters augmented by the spectral variable $z$ in the
    rational Lax matrix $L(z)$. The extended system therefore does not preserve  the symplectic foliation,
since it has a transversal component corresponding to the ``explicit'' dependence on the deformation parameters.
    However, the {\it transverse foliation} may be viewed as generated by a locally free abelian group action that preserves the
    Poisson structure, defining a local isomorphism between  neighbouring leaves of the symplectic foliation.
    The quotient by this group action may be identified with any of the neighbouring symplectic leaves (augmented by
    some trivial Casimir elements, consisting of the {\em exponents of formal monodromy} \cite{BJL, JMU}), and the
    projection of the infinitesimal isomonodromic deformation vector field to this quotient is the corresponding Hamiltonian  vector field.
        
     The key problem that needs to be resolved in the general case,   with  irregular singularities of arbitrary Poincar\'e rank, is:      {\em  how do  we define these ``explicit derivative'' vector fields, and how do we choose the spectral invariant Hamiltonians  whose vector fields, when added to  the ``explicit derivative"  ones, generate the corresponding $1$-parameter family of isomonodromic deformations?}
     
    In the  cases studied earlier \cite{JMU,  JM, HTW, H2,  HR}, this was very straightforward.  For the Schlesinger systems,   the ``explicit parameters'' are simply  the loci of the first order poles of the Lax matrix governing the Schlesinger equations. For systems with one further non-Fuchsian singularity at $\infty$, with Poincar\'e index $1$, the additional deformation parameters are the eigenvalues  of the  Lax matrix at $z=\infty$; i.e., the constant matrix term added to the Schlesinger Lax matrix  \cite{H2}. Of the six Painlev\'e equations, two ($P_{VI}$ and $P_V$ respectively) are special cases of these, and for the remaining four ($P_I  - P_{IV})$, the additional deformation parameter is one of the higher Birkhoff  invariants identified in \cite{JM} and used in the canonical  parametrization \cite{JM, HR} of their $2\times 2$ Lax matrices.
     
    In this work, the problem is solved for  arbitrary (nonresonant) rational isomonodromic deformation
    equations of the type introduced in  \cite{JMU, JM}, placing these in the Hamiltonian framework  provided 
    by the rational $R$-matrix structure.  Some of the results presented here have appeared earlier  \cite{H2, H4, BertoMo} in partial form.  In \cite{MaMo}, the $P_{II}$ hierarchy, which consists of  (reduced) isomonodromic deformation systems involving $2 \times 2$ polynomial Lax matrices of any degree, plus a first order pole at $z=0$ was treated using the same rational $R$-matrix Poisson bracket structure. 
 In \cite{GMR}, the Hamiltonian structure and quantization of rational isomonodromic deformations were studied, 
with irregular singularities obtained via confuence of poles.
 After this work was completed, we learned of ref.~\cite{Yam},  in which some of the earlier results of \cite{H2, H4, BertoMo},  which are included here in Section \ref{spectralinvars}, were re-derived and extended in ways that overlap with parts of Sections \ref{Birkhoff_conn_commut_Poisson} and \ref{birkhoff_fibration}. 
Another recent work \cite{MOA} derived a Hamiltonian representation of rational $2 \times 2$ isomonodromic systems in a different way, making use of the {\it spectral Darboux coordinates} introduced in \cite{AHH2}. 
      Our purpose here  is to present a complete, self-contained account, which contains all results known to date.    
  
   The next three subsections recall the main implications of classical $R$-matrix theory,
    and summarize how the isospectral equations generated by a suitably chosen class
   of spectral invariant Hamiltonians, which are  {\em dual}  to the deformation parameters, in a natural sense,
     are converted into the isomonodromic deformation equations under consideration.
%%%%%%%%%%%%%%%  Subsection 1.2  Rational $R$-matrix structure and isospectral systems %%%%%%%%%%%%%%
\subsection{Rational $R$-matrix structure and isospectral systems}
\label{rational_R_matrix_isospectral}

The phase space $\LL_{r, \db}$ considered throughout this work consists of  complex, traceless $r\times r$ rational 
matrix-valued functions of a spectral variable $z\in\mathbb P^1$
\be
L(z)  =  -\sum_{j=0}^{d_\infty-1} L^{\infty}_{j+2} z^j +\sum_{\nu=1}^N
\sum_{j=1}^{d_\nu+1} \frac {L^\nu_{j}}{(z-c_\nu)^j},
\label{rational_Lax_matrix}
\ee
henceforth referred to as the {\em Lax matrix}, where $\db := (d_1, \dots, d_N, d_\infty) \in \Nbb_+^{N+1}$.  
The matrix differential $L(z)d z $ has pole divisor
bounded by
\be
{\rm div}_{pole}\le(L(z)d z\ri) \geq -(d_\infty+1) \infty - \sum_{\nu=1}^N (d_\nu+1) c_\nu,
\label{L_max_pole_deg}
\ee 
and the finite pole loci $\cb=(c_1, \dots, c_N)$ are distinct.
The leading coefficients $\{L^\nu_{d_\nu+1}\}_{\nu=1,\dots, N, \infty}$  of the polar parts, both at the finite $c_\nu$'s  and at $c_\infty=\infty$, 
are assumed diagonalizable,  with distinct eigenvalues (the {\em nonresonant} condition \cite{JMU, BertoMo}). 
By conjugation with a constant matrix we may, without loss of generality, choose $L^{\infty}_{d_\infty+1}$ to be  diagonal, with the entries all 
conserved quantities. The set  $\LL_{r, \db}$ of such $L(z)$'s may be interpreted as a Poisson submanifold of the dual space $L^*_R\grgl(r)$ of the modified loop algebra 
$L_R\grgl(r)$ (viewed as smooth maps $L:S^1 \ra \grgl(r)$ from the unit circle $S^1 = \{z \in \Cbb  \vert |z|=1\}$)
with respect to the modified Lie-Poisson bracket structure corresponding to the so-called {\it split} 
rational classical $R$-matrix \cite{RS1, Sem, RS2}.

Splitting the space $L_R\grgl(r) \sim L\grgl(r)$ as a direct sum 
\be
L_R\grgl(r) = L_+\grgl(r) \oplus L_-\grgl(r)
\ee
of subspaces (and subalgebras) consisting of elements $X_+ \in L_+\grgl(r)$ that 
 admit analytic continuation inside the unit circle (or, simply, positive power Fourier series)
\be
X_+ = \sum_{i=0}^\infty X_iz^i,\ X_i \in \grgl(r), \ z=e^{i\theta} \in S^1
\ee
and  those $X_- \in L_-\grgl(r)$  that admit analytic continuation outside, 
with $X_-(\infty)=0$ (or negative power Fourier series )
\be
X_-= \sum_{i=1}^\infty X_i z^{-i},\ X_i \in \grgl(r), \ z=e^{i\theta} \in S^1,
\ee
the modified Lie bracket, denoted $[X, Y]_R$ is defined  by
\be
[X_+, Y_+]_R = [X_+, Y_+], \quad [X_-, Y_-]_R = -[X_-, Y_-], \quad [X_+, Y_-]_R = 0.
\ee

The dual space  $L_R^*\grgl(r)$ (or  $L^*\grgl(r)$)  is identified  with $L_R\grgl(r)$ (or $L\grgl(r)$) through the pairing
\be
\mu(X) = \frac{1}{2\pi i} \oint_{z \in\gamma} \tr\left(\mu(z) X(z)\right) d z ,    \quad \mu \in L^*\grgl(r), \ X \in L\grgl(r).
\label{trace_res_pairing}
\ee
The annihilators $(L_\pm\grgl(r) )^0 \ss L_R^*\grgl(r)$ of the subalgebras $L_+\grgl(r) $ and $L_+\grgl(r) $ are therefore identified as 
\be
(L_+\grgl(r) )^0 = L_+\grgl(r) = (L_-\grgl(r) )^* , \quad (L_-\grgl(r) )^0 = L_-\grgl(r) = (L_+\grgl(r) )^* 
\ee
and these are mutually disjoint Poisson subspaces of $L^*_R\grgl(r)$.
Here $\gamma$ can be chosen as the unit circle $S^1$ centered at the origin  oriented counterclockwise, 
but it will be convenient, when considering finite dimensional Poisson subspaces of  $L^*_R\grgl(r)$ consisting
 of rational elements $L\in L^*_R\grgl(r)$ of the form (\ref{rational_Lax_matrix}), to  use the same notational conventions,  
but  replace $S^1$ by a circle of any radius, chosen so that all the finite poles lie in the interior
 of the integration contour   $\gamma$.

The canonical Lie-Poisson bracket with respect to the modified Lie algebra structure $L_R\grgl(r)$ 
\be
\{f,\, g\}_R \big|_{\mu \in L^*\grgl(r)} := \mu\big([d f ,  d g]_R\big) = \mu\big([(d f,)_+, \, (d g)_+] - [(d f)_-, \, (d g)_-]\big )  , \quad f, g \in C^1(L^*\grgl(r)),
\label{split_R_loop_Lie_PB}
\ee
is dual to the Lie bracket on $L_R\grgl(r)$, and has a multitude of finite dimensional Poisson subspaces; 
in particular, the space $L^*_{(\cb, \db)}\grgl(r)$ of  rational matrices of the form (\ref{rational_Lax_matrix}) for fixed  pole loci 
\be
(\cb, \infty), \quad \cb:= \{ c_1, \dots, c_N\}
\label{pole_loci}
\ee
and maximal degrees
\be
 \db:=\{d_1, \dots, d_N, d_\infty\}.
 \label{max_deg_d}
\ee
This may equivalently be identified with the Lie-Poisson space consisting of the dual space 
to the finite dimensional Lie algebra
\be
L_{\db}\grgl(r) := \oplus_{\nu=1}^N \grgl^{(d_\nu )}(r)  \oplus  \grgl^{(d_\infty +1 )}(r),
\ee
where $\grgl^{(j)}(r)$ denotes the $j$th jet extension of $\grgl(r)$, and the leading polynomial coefficient 
matrix $L^{(d_\infty)}$, whose entries consists only of Casimirs elements, is chosen as diagonal.
The corresponding Lie algebra $L_{(\cb, \db)}\grgl(r)$ may be viewed as the quotient of the full loop algebra $L\grgl(r)$ by the ideal
\be
\label{Ideal}
\grI_{\cb, \db}:=z^{-d_\infty-1} \prod_{\nu=1}^N( z-c_\nu)^{d_\nu +1} L_R\grgl(r) \ss L_R\grgl(r), 
\ee
in which the pole loci $(c_1, \dots, c_n)$ are ``spectators'', their values 
giving different equivalent identifications of  $L_{(\cb, \db)}\grgl(r)$, for varying $\cb$,
 as quotients of $L_R\grgl(r)$ by equivalent ideals.
Taking a disjoint union over these, the pole locations $\{c_1, \dots, c_N\}$,  may be viewed as further coordinates, which are Casimir 
functions on the larger Poisson subspace, consisting of the product
\be
( \Cbb^N)' \times L^*_{\db}\grgl(r),\quad (\Cbb^N)' := \{(c_1, \dots, c_N)\in \Cbb^N \ \vert \ c_i \neq c_j \ \text{for} \ i\neq j\},
\ee
 that Poisson commute  amongst themselves and with the elements of  $L_{\db}\grgl(r)$.

The coadjoint action of the corresponding split loop group,
\be
L_+\grGl(r) \times L_-\grGl(r) = \{(g_+(z), g_-(z))\}
\ee
where $g_-(\infty) = \Ib$, is given by {\em dressing transformations}:
\be
Ad_R^*\left(g_+, g_-\right): (X_+ + X_-) \mapsto (g_- X_+ (g_-)^{-1})_+ + ((g_+ )^{-1}X_- g_+)_-,
\label{dressing_transf}
\ee
where $(\dots)_\pm$ denotes projection to the positive and negative parts of the Fourier series.
The orbits under this action are the symplectic leaves of the $R$-matrix Poisson structure $\{ \, , \, \}_R$,
which we henceforth just denote as $\{ \, , \, \}$.

Written  in terms of matrix elements  of $L$ evaluated at two values $(z, w)$ of the loop  parameter,
viewed as linear functionals on the loop algebra,  the Lie-Poisson brackets corresponding to the  
split classical rational $R$-matrix structure  (\ref{split_R_loop_Lie_PB}) are given by
\be
\{L_{ab}(z), L_{cd}(w) \}= \frac{1}{z-w} \Big( (L_{ad}(z) -L_{ad}(w) )\delta_{cb}  -  (L_{cb}(z) -L_{cb}(w) )\delta_{ad}  \Big).
\label{split_rational_Rmatrix_PB}
\ee
Classical $R$-matrix theory \cite{RS1, Sem, RS2} then implies that:
\begin{enumerate}
\item All elements of the ring $\II^{\Ad^*}(L^*\grgl(r))$ of
(unmodified) $\Ad^*$ invariant functions of $L(z)$ (i.e., the ring of spectral invariants) Poisson commute amongst themselves.
\be
\{f, g\} =0, \quad \forall \ f, g\in \II^{\Ad^*}(L^*\grgl(r)).
\ee
This means that, on $L_{r, \db}$,  all elements of the ring $\II^{\Ad^*}(L^*\grgl(r))$ generated by the coefficients of the characteristic polynomial
\be
\det(L(z) - \lambda \Ib)=0
\label{char_eq_L}
\ee
defining the (planar) spectral curve $\CC_0$ Poisson commute.
\item
The Hamiltonian vector field $\Xb_H$  generated by any element $H\in \II^{\Ad^*}(L^*\grgl(r))$ is given by a commutator 
\bea
\Xb_H(X)&\& =\{X, H \} = [R_s(d H), X], 
 \label{Hamvec_commut}\\
 \forall \ H &\&\in \II^{\Ad^*}(L^*\grgl(r)), \ X \in L\grgl(r),
 \nonumber
\eea
where $X \in L\grgl(r)$ is viewed as a linear functional on $L^*\grgl(r)$ under the 
pairing (\ref{trace_res_pairing}) and $R_s$ is the endomorphism of $L\grgl(r)$ defined by
\be
R_s(Y_+ + Y_-) = s Y_+  + (s - 1) Y_-,    \quad Y \in L\grgl(r)
\ee
for any $s \in  \Cbb$. In particular,  
\be
R_1(Y_+ + Y_-) = Y_+, \quad  \text{and} \quad R_{0}(Y_+ + Y_-)= - Y_-.
\label{Rmatrix}
\ee
\begin{remark}
For the isospectral systems generated by such Hamiltonians, the choice of $s\in \Cbb$ is irrelevant,
since changing it only adds a term proportional to $dH(L)$, which is in the commutant of $L$.
But when the system is modified, as in Sections \ref{examples_nonauton_isomon} and \ref{rational_isomon_ham_nonauton},
by the addition of a transverse ``explicit derivative'' vector field, only one specific choice of $s$, usually $s=0$ or $1$, 
renders the system isomonodromic for any of the relevant spectral invariant Hamiltonians. 
\end{remark}
\item
The Hamiltonian flow generated by any element  $H\in \II^{\Ad^*}(L^*\grgl(r))$
of the spectral ring leaves invariant the spectral curve (\ref{char_eq_L}), and
hence the flow is isospectral. The time dependence of the corresponding integral
curves $L(z,t)$ is determined by the Lax equation
\be
\frac{d L}{d t  } = [R_s(d H), L],
\label {Lax_eq}
\ee
and the Hamiltonian flow is given by
\be
f_H(t): L(0) \ra L(t) =\Ad^*_R\left(g_+(t),g_-(t)\right) (L(0)) , 
\ee
where $(g_+(t), g_-(t))$   is determined as the solution of the Riemann-Hilbert factorization problem
\be
g_+(t) g_-(t) = e^{t d H(L(0))}, \quad  (g_+(t), g_-(t))\in L_+\grGl(r) \times L_-\grGl(r).
\ee
\end{enumerate}

%%%%%%%%%%%%%%%  Subsection 1.3  Nonautonomous deformations: isomonodrmic systems %%%%%%%%%%%%%%
\subsection{Examples of nonautonomous deformations as isomonodromic systems}
\label{examples_nonauton_isomon}

The study of isospectral Hamiltonian systems  generated by $Ad^*$ invariant functions within the rational classical $R$-matrix 
Poisson bracket structure on (the dual space of) loop algebras  was developed in \cite{AHH1, AHH2, H1}
 for  autonomous  systems and  this was extended  to nonautonomous isomonodromic ones in \cite{HTW,  H2,  HR, H4}. 
 The simplest case consists of the Schlesinger equations,  which generate isomonodromic deformations of the Fuchsian system
\be
\frac{ \partial \Psi(z)}{\partial z} =  L^{\text{Sch}}(z)\Psi(z),  \quad \Psi(z)\in {\grGl}(r),
\label{schlesinger_z_deriv_eq}
\ee
where
\be
L^{Sch}(z) := \sum_{\nu =1}^N \frac{L^{\nu}} {z- c_\nu}.
\label{fuchsian_Lax_matrix}
\ee
The Casimir elements that serve as deformation parameters are  the loci  $(c_1, \dots, c_N)$ of the poles.
The corresponding spectral invariant Hamiltonians are
\be
H_\nu := \frac{1}{2}\res_{z= c_\nu} \tr \left(L^{Sch}\right)^2d z, \quad \nu =1,\dots, N
\label{Schelsinger_hams}
\ee
and the Hamiltonian vector fields, acting on $L^{Sch}(z) $, are given by commutators with the matrices
\be
R_0(d H_\nu) = -  \left( d H_\nu\right)_- = - \frac{L^\nu}{z-c_\nu} .
\ee
The resulting nonautonomous  equations, in which the explicit dependence of the Lax matrix
$L(z)$ on the pole loci $\cb$ is taken into account, are thus
\be
\frac{\partial L^{\text{Sch}}(z)}{\partial c_\nu} = \left[ -\frac{L^\nu}{z - c_\nu}, L^{\text{Sch}}(z)\right] + \frac{L^\nu}{(z-c_\nu)^2}, \quad \nu=1, \dots, N
\label{schlesinger_ham_eqs}
\ee 
The additional term $\frac{L^\nu}{(z-c_\nu)^2}$ in eq.~(\ref{schlesinger_ham_eqs})  must be included because 
of the explicit dependence of $L(z)$ on the pole locations $\{c_\nu\}_{\nu =1, \dots, N}$.
By equating the residues, these are equivalent to the Schlesinger equations
\begin{subequations}
 \bea
\frac{\partial L^{\mu}}{\partial c_\nu} &\& = \frac{[L^{\mu}, L^\nu]}{c_\mu - c_\nu}, \quad  \forall \   \nu \neq  \mu, 
\label{schlesinger_eqs_mu_nu}
 \\
\frac{\partial L^{\mu}}{\partial c_\mu} &\& = - \sum_{\nu=1, \ \mu \neq \nu}^N\frac{[L^{\mu}, L^\nu]}{c_\mu - c_\nu}.
\label{schlesinger_eqs_mu_mu}
\eea
\label{schlesinger_eqs}
\end{subequations}
{\hskip -8 pt} They are also  the compatibility conditions for the overdetermined system consisting of (\ref{schlesinger_z_deriv_eq}),
together with the infinitesimal deformation equations
\be
\frac{ \partial \Psi}{\partial c_\nu} =   -\frac{L^{\nu}}{z- c_\nu} \Psi, \quad \nu =1, \dots, N,
\label{schlesinger_deform_eqs} 
\ee
and hence imply  invariance of the monodromy of the operator 
\be
\DD^{L^{sch}}:= \frac{\partial}{\partial z} - L^{Sch}
\ee
under changes in the pole locations.

Viewed geometrically, eqs.~(\ref{schlesinger_z_deriv_eq}),  (\ref{schlesinger_deform_eqs})  represent parallel transport with respect to the connection form 
\be
\Omega:=-L^{\text{Sch}}d z + \sum_{\nu=1}^N\frac{ L^{\nu}}{z- c_\nu} d c_\nu
\ee
over the product of the spectral parameter space $\{z\in \Pbb^1\}$ ) and the space of distinct pole loci  $\{\cb \}$, with the loci of poles removed.
Their compatibility conditions are equivalent to  the commutativity of the covariant derivatives over the parameter space 
\begin{subequations}
\bea
\left[\frac{\partial}{\partial z} - L^{Sch}(z),  \frac{\partial}{\partial c_\nu} + \frac{L^\nu}{z-c_\nu}\right] &\& =0,
\label{schlesinger_zero_curv_z_nu}
 \\
&\& \cr
\left[\frac{\partial}{\partial c_\mu} + \frac{L^\mu}{z-c_\mu}, \frac{\partial}{\partial c_\nu} + \frac{L^\nu}{z-c_\nu}\right] &\& =0,
\quad 1 \leq  \mu, \nu \leq N. 
\label{schlesinger_zero_curv_mu_nu}
\eea
\label{schlesinger_zero_curv}
\end{subequations}
{\hskip -4 pt}These are equivalent to the Schlesinger equations (\ref{schlesinger_eqs_mu_nu}), (\ref{schlesinger_eqs_mu_mu}),
which therefore are interpretable as zero curvature equations for a flat connection.
The additional term $\frac{L_\nu}{(z-c_\nu)^2}$ in eq.~(\ref{schlesinger_ham_eqs})  turns 
the isospectral Hamiltonian equations into the zero curvature equations (\ref{schlesinger_zero_curv}) 
 because of the identity\footnote{The notation $\frac{\partial^0}{\partial c^0_\nu} $ for the ``explicit derivatives'' will
 be changed to $\nabla_{c_\nu}$ in what follows 
 and, more generally, $\nabla_t := \frac{\partial^0}{\partial t^0}$ for the further deformation parameters $\{t=t^\nu_{ja}\}$ 
 to be introduced  in (\ref{princ_part_coeffs_nu}),  (\ref{princ_part_coeffs_inf})  below.}
 \be
  \frac{\partial} {\partial z} \left(-\frac{L^\nu}{z-c_\nu}\right)   = \frac{\partial^0 L^{Sch}}{\partial c^0_\nu}
  = \frac{L^\nu}{(z-c_\nu)^2}, 
  \label{explicit_cross_deriv}
\ee
 which we refer to as an {\it isomonodromic identity}, where $\frac{\partial^0 L^{Sch}}{\partial c^0_\nu}  $ 
  denotes the {\it explicit derivative} with respect to the pole
location parameter $c_\nu$, and the entries of the matrices $\{L^\nu\}_{\nu=1, \dots, N}$ are considered as independent coordinates. 
This simple identity is the essential reason why the $R$-matrix approach which, in the autonomous case gives
 isospectral (Lax) equations, when applied to the nonautonomous system obtained by identifying the deformations parameters
as pole loci, gives rise to the zero curvature equations (\ref{schlesinger_zero_curv}).

Note that there are two ingredients leading to this result. 
The first is that the explicit parametric dependence in the nonautonomous system is just
through the location of the poles, which are Casimir elements of the Poisson structure.
The second is that the exact choice (\ref{Schelsinger_hams}) of the corresponding Hamiltonians from amongst 
the various possible spectral invariants implies the  isomonodromic  identity (\ref{explicit_cross_deriv}).
These must be matched in a special {\it dual} way with the deformation parameters
in order that the identity (\ref{explicit_cross_deriv}), resulting in a zero 
curvature system, be satisfied. The  notion of {\it explicit dependence} and {\it dual pairing} of the deformation
parameters with the Hamiltonians, which is clear in this special case, is not obvious in the more general
case of arbitrary rational Lax matrices. Making this precise will be one of the main points of the 
subsequent development.  (See Theorems \ref{thmcasi} - \ref{zero_curv_Lax}).
 
A slightly more general case was  treated similarly  in \cite{H2} where, in addition to the first order poles appearing 
in (\ref{fuchsian_Lax_matrix}), a constant diagonal matrix $B= \diag(b_1, \dots, b_r)$ was added to the Lax matrix, giving
\be
L(z) = B + L^{Sch}(z),
\label{schlesinger_Lax_plus_B}
\ee
and hence adding a second order pole in the connection form $L(z)d z $  at $\infty$. The deformation parameters 
 $(b_1, \dots, b_r)$ are again Casimir elements in the  $R$-matrix Lie-Poisson structure and there
 is a corresponding special set of $r$ spectral invariant Hamiltonians $\{K_a\}_{a=1, \dots, r}$  (see \cite{H2}) generating the 
Hamiltonian vector fields as commutators.  The deformation matrices $\{(d K_a)_+\}$ again satisfy the 
necessary equality 
\be
\frac{\partial (dK_a)_+}{\partial z} = \frac{\partial^0 L}{\partial b^0_a}  = E_{aa}, \quad a=1, \dots, r
\label{B_isom_ident}
\ee
between their derivatives with respect to the spectral parameter $z$  and
the ``explicit'' derivatives of the Lax matrix with respect to the parameters $(b_1, \dots, b_r)$
(where $E_{ab} \in \grgl(r)$ is the elementary matrix whose only nonvanishing entry  is a $1$ in the $(a,b)$ position).
 Starting with the isospectral  Hamiltonian Lax equations following from the $R$-matrix structure, adding the  explicit derivatives
with respect to the nonautonomous deformation parameters $(b_1, \dots, b_r)$ in the Lax matrix and using
the {\em isomonodromic identity} (\ref{B_isom_ident})  again assures that the isospectral equations appearing
in the autonomous case become zero curvature ones in the nonautonomous one. The
compatibility conditions of the deformation equations again imply the invariance of the (generalized) monodromy.

In \cite{HR} the rational $R$-matrix approach was also applied to deriving the
five Painlev\'e transcendent equations $P_I$ - $P_{V}$, which are all reductions of
$2 \times 2$ rational isomonodromic deformation equations, with pole divisor having total degree $-4$.
(The $P_{VI}$ case is just a symmetry reduction of the  $r=2$ Schlesinger system with $3$ finite simple poles, plus one at $\infty$.)
In \cite{HTW}, it was applied to rational systems in $L_R^*\grgl(2)$  with an arbitrary number of first order
poles at the finite points $\{c_\nu\}_{\nu=1, \dots, N}$ plus an additional irregular singularity of Poincar\'e index $2$ at $\infty$.

The question that naturally occurs is whether this approach can be extended 
to the general class of rational isomonodromic deformation systems introduced in \cite{JMU, JM}, in 
which the Lax matrix is of the form (\ref{Schelsinger_hams}), and the connection has any number of irregular 
singularities of arbitrary Poincar\'e\ rank at finite points or at $z=\infty$.
Completing the analysis for the general case is the main purpose of the present work. 
It follows along lines similar to the cases previously treated, but gives a clearer notion of what is
meant by the {\it explicit derivative} of the Lax matrix with respect to the further deformation parameters 
which, as before, are viewed as functions on the phase space. As in the previous cases, 
 it turns out that only Casimir elements can play this r\^ole and, in fact,  (nearly) all of them do.  
 
 The full set of these provide a transversal,  regular foliation complementary to the  one  
 given by the symplectic leaves, which are the ``dressing transformation'' orbits under the $R$-matrix Lie algebra structure.
 In addition to the pole loci $\{c_\nu\}_{\nu = 1, \dots, N}$,  the further Casimir elements  that serve as deformation parameters turn out to coincide with the 
higher {\it Birkhoff invariants}  $\{t^\nu_{ja}\}_{\nu=1, \dots, N, \infty,\, j=1, \dots, d_\nu, \, a=1, \dots, r}$ appearing in  \cite{BJL, JMU, JM}, which characterize the formal asymptotic behaviour of a fundamental system
 near the irregular singular points, but may also be expressed as spectral invariant functions  of the Lax matrices, as in eqs.~(\ref{princ_part_coeffs}) 
 
 The second ingredient consists of finding the correct spectral invariant Hamiltonians that are paired ``dually'' with these Casimir elements. 
 This turns out to have an elegant solution in terms of the structure  of the spectral curve, and the local singularity structure of the 
 naturally associated meromorphic differential  over the poles  in the Lax matrix. (See eqs.~(\ref{spectral_Isomon_hamiltonians_t_nu}), (\ref{spectral_Isomon_hamiltonians_t_inf}), (\ref{SchlesHam}).) The ``isomonodromic identities''  (\ref{isomon_cond})  that allow the associated  isospectral systems 
  given by the $R$-matrix dynamics to be converted into isomonodromic ones are ``reverse engineered'', in the sense
 that eqs.~(\ref{isomon_cond}) are used as the {\it definition} of  what the ``explicit derivatives''  mean. 
It is then {\em verified}  (Theorem \ref{rational_isomon_systems}) that these can, indeed,  be interpreted as a set of commuting vector 
fields on the phase space.  By Frobenius' theorem,  they may be interpreted as commuting directional derivatives along transversal curves defined 
by fixing all but one of the Birkhoff invariants, viewed as coordinate functions on the phase space. They furthermore preserve the 
Poisson structure, and therefore provide an integrable distribution transverse to the symplectic foliation, allowing at least a
local identification of the neighbouring symplectic leaves in a tubular neighbourhood as the orbit of an abelian group action
on a single one of these. This gives a consistent Hamiltonian framework for the deautonomization of the isospectral equations  
in the general case,  as summarized in the following subsection.

%%%%%%%%%%%%%%%  Subsection 1.4  Isomonodromic systems as Hamiltonian nonautonomous deformations %%%%%%%%%%%%%%
\subsection{Rational isomonodromic systems and nonautonomous Hamiltonian deformations}
\label{rational_isomon_ham_nonauton}

In general, we consider isomonodromic deformations of rational covariant derivative operators:
\be
\DD_z^L := \frac{\pa}{\pa z} - L(z),  
\label{rational_covar_deriv}
\ee
where $L(z)$ is a rational Lax matrix of the form (\ref{rational_Lax_matrix}), 
with fundamental systems $\Psi(z)$ of solutions of the equation
\be
 \frac{\pa \Psi(z)}{\pa z} = L(z) \Psi(z), \quad \Psi(z) \in \grGl(r). 
 \label{rational_cov_deriv_eq}
\ee

In addition to the pole loci $\{c_\nu\}_{\nu=1, \dots, N}$, the independent variables parametrizing the deformations  will be identified 
with a subset of the Casimir elements, defined in  eqs.~\eqref{princ_part_coeffs_nu}, \eqref{princ_part_coeffs_inf}, 
denoted $\{t^\nu_{j a}, t^{\infty}_{j a}\}$, which will be shown to coincide with the {\em higher Birkhoff invariants} 
\cite{BJL} of Definition  \ref{birkhoff_invar}. Here, as in (\ref{rational_Lax_matrix}),  the indices $\nu = 1, \dots, N$ denote the 
finite pole locations $\{z=c_\nu\}$, with corresponding negative powers  $j=1, \dots , d_\nu $  in the principal part of 
$L(z)$ at $z=c_\nu$ and $j=  1, \dots  d_\infty$  at $z=\infty$,  the positive powers in the polynomial part,
The indices  $a=1, \dots , r$ correspond to the $r$ different local solutions $\{\lambda_a(z)\}_{a=1, \dots, r}$  of the characteristic equation (\ref{char_eq_L}) 
near  the poles $\{z=c_\nu\}_{\nu=1, \dots, N, \infty}$, given as Laurent series in a punctured neighbourhood of each pole location.
 Equivalently, the $a$'s may be viewed as indexing the sheets of the 
spectral curve $\CC$ obtained by compactifying the planar curve $\CC_0$ defined by the characteristic equation \eqref{char_eq_L}.
Because of the assumption that the leading coefficients around each singularity of $L$ have distinct eigenvalues, given by a meromorphic 
function $\lambda$ on $\CC$  satisfying the characteristic equation (\ref{char_eq_L}), near each of the points $\{p_\nu^{(a)}, \infty^{(a)} \in \mathcal C\}_{a=1,\dots, r,}$ 
over the points $\{z=c_\nu\}_{\nu=1, \dots, N, \infty}$, this has $r$ distinct local Laurent series' solutions$\{\lambda_a(z)\}_{a=1, \dots, r}$.\footnote{The various local Laurent series $\{\lambda_a(z)\}$ near $\{z=c_\nu\}_{\nu=1, \dots, N, \infty}$ 
depend, of course, on $\nu$ as well,  but we omit  indicating this explicitly to avoid a plethora of indices.}

In the autonomous case, the meromorphic differential $\lambda \, d z$  on the spectral curve $\CC$
plays a fundamental r\^ole  in  explicitly integrating the Hamiltonian Lax equations corresponding to elements $H\in\II^{(\cb, \db)}$  
of the spectral ring generated by the coefficients of the eigenvector equation (\ref{char_eq_L}) in terms of abelian functions  \cite{AHH1, AHH2, H1}.
A complete set of generators of the center of the Lie-Poisson algebra defined above (i.e.  the Casimir elements)
is given by the loci $\{c_\nu\}_{\nu =1, \dots, N}$ of the finite poles, together with the following further spectral invariants
\begin{subequations}
\bea
t^\nu_{j a}&\& :=- \res_{z=c_\nu } (z-c_\nu)^j \lambda_a(z) d z ,
\label{princ_part_coeffs_nu}
\\
\quad \nu &\&=1, \dots, N, \quad  j =0, \dots d_\nu,  \quad a=1, \dots, r,
\cr
 t^\infty_{j a} &\&:= -\res_{z=\infty } z^{-j} \, \lambda_a(z) d z , 
 \label{princ_part_coeffs_inf}
 \\
  j &\&=1, \dots d_\infty, \quad a=1, \dots, r, 
  \nonumber
\eea
\label{princ_part_coeffs}
\end{subequations}
{\hskip -10 pt} which are just the coefficients of the principal parts of the Laurent expansion of $\lambda_a(z)\, d z  $ at the pole locations.
As shown in Section \ref{secMonodromy},  these may be identified with the {\em Birkhoff invariants} (Definition \ref{birkhoff_invar})
defining the formal local asymptotics of a fundamental system of solutions of  (\ref{rational_cov_deriv_eq})
in a neighbourhood of any finite  irregular singular point $z=c_\nu$,  $(d_\nu > 0)$ and at $z=\infty$.
The parameters  
\be
t^\nu_{0a}:= - \res_{z=c_\nu }  \lambda_a(z) d z,\ \  \nu=1,\dots, N, \quad a=1, \dots, r,
\label{form_mon_nu}
\ee
are known as the {\em exponents of formal monodromy} \cite{JMU, JM}).

Dual to these are the following non-Casimir spectral invariants
\begin{subequations}
\bea
H_{t^\nu_{j a}} &\&:= - \res_{z=c_\nu}\frac 1{j(z-c_\nu)^{j}}\, \lambda_a(z)d z , 
\label{spectral_Isomon_hamiltonians_t_nu} \\
\nu  &\&=1, \dots, N,  \quad   j=1, \dots d_\nu,  \quad a=1, \dots, r,
\cr
&\& \cr
 H_{t^\infty_{j a} }: &\&=  -\res_{z=\infty}\frac{z^j}j\, \lambda_a(z) d z ,
 \label{spectral_Isomon_hamiltonians_t_inf}  \\
  &\&j=1, \dots d_\infty, \quad  a=1, \dots r, 
  \nonumber
\eea
 \label{spectral_Isomon_hamiltonians_t}  
\end{subequations}
{\hskip -6 pt} which are the ``mirror image'' coefficients  of the principal parts of the local Laurent expansion  
of $\lambda_a(z)dz$ near $z=c_\nu$, consisting of the coefficients of the first $d_\nu$ positive powers of $z-c_\nu$ in the analytic part,
and the inverse powers at $z=\infty$, and
\be
H_{c_\nu} := \frac{1}{2} \res_{z=c_\nu} \tr( L^{2}(z))d z.
\label{SchlesHam}
\ee
(Note that there are no non-Casimir spectral invariants dual to the exponents of formal monodromy $\{t^\nu_{0a}\}_{\nu=1, \dots, N, \, a= 1 , \dots r}$.)
In addition to these, we also have the exponents of formal monodromy at $z=\infty$ which we denote as
\be
 H^\infty_a := t^\infty_{0a} = - \res_{z=\infty}  \lambda_a(z) d z,\ \  a=1, \dots, r.
\label{form_mon_infty}
\ee
Unlike those $\{ t^\nu_{0a}\}_{\nu=1, \dots, N, a=1, \dots , r}$ at the finite poles $z=c_\nu$,
these are not Casimir elements, but  dynamical spectral invariant Hamiltonians  generating nontrivial Hamiltonian flows, 
whose Hamiltonian vector fields are given by the commutators 
\be
\Xb_{H^\infty_a}L(z)= \{L(z), H^\infty_a\} = [E_{aa}, L].
\label{HamVec_inf_a}
\ee
The flows they generate, consisting of conjugation by invertible, $z$ independent diagonal matrices,
are therefore {\em both} isospectral and isomonodromic, and should be understood as symmetries
of the rational isomonodromic deformation systems of \cite{JMU, JM}.

It is important to note that, although  $\{t^\nu_{j  a}, t^\infty_{j  a}, H_{t^\nu_{j  a}}, H_{t^\infty_{j  a}},  H_{c_\nu}, H^\infty_a \}$
are defined here by evaluation of residues of moments of $\{\lambda_a(z) \, d z \}_{a=1, \dots, r} $ at the
poles $\{z=c_\nu\}_{\nu=1, \dots, N, \infty}$,  they may also be computed explicitly as polynomial expressions in the
entries of the matrix terms $\{L^\nu_j\}$ appearing in eq.~(\ref{rational_Lax_matrix}), and  (see Corollary \ref{corratl}) 
depend {\it rationally} on the differences of the eigenvalues of the leading coefficients $L^\nu_{d_\nu+1}$. 

To each of the Hamiltonians 
$\{H_{t^\nu_{ja}},H_{c_\nu}\}_{\nu=1, \dots, N, \infty,  \, j=1, \dots d_\nu, \, a=1, \dots , r}$ 
defined above, there corresponds a Hamiltonian vector field  
$\{\Xb_{H_{t^\nu_{ja}}}, \Xb_{H_{c_\nu}} \} $ defined by the equations
\begin{subequations}
\bea
\Xb_{H_{t^\nu_{ja}}}L(z)&\& = \le\{L(z), H_{t^\nu_{ja}}\ri\} =\Big[U^\nu_{ja} , L(z)\Big] , 
\label{HamVec_t}
 \\
\Xb_{H_{c^\nu}} L(z) &\& = \le\{L(z), H_{c_\nu}\ri\}= \Big[V^\nu , L(z) \Big] ,
\label{HamVec_c}
\eea
\end{subequations}
where the Poisson bracket \eqref{split_rational_Rmatrix_PB} is applied to each of the entries of $L(z)$.  
Recall \cite{JMU, JM} that the  generalized {\it isomonodromic deformations} are determined by a system of PDEs of the  form
\begin{subequations} 
\bea
\frac{\partial \Psi(z)}{\partial t^\nu_{j a}} &\&= U^\nu_{j a}(z) \Psi(z), \quad \nu= 1,\dots, N,\infty, \quad j=1, \dots, d_\nu, \quad a =1, \dots,r ,
\label{isomonPDE_U}  \\
\frac{\partial  \Psi(z)}{\partial c_\nu} &\&= V^\nu(z)\Psi(z),\qquad  \nu=1,\dots, N,
\label{isomonPDE_V}
\eea
\label{isomonPDE_UV} 
\end{subequations}
{\hskip -8 pt} where the matrices $\{U^\nu_{ja}(z), V^\nu(z)\}$ (whose definition is given in Section \ref{isomon_deform}, eqs.~(\ref{defU}), (\ref{defV}))
 depend rationally on $z$ and are uniquely determined in terms of the entries of $L(z)$ (see Theorem \ref{zero_curv_Lax}). These equations generate deformations such that the {\it extended monodromy data} of the ODE \eqref{rational_cov_deriv_eq} (monodromy matrices, connection matrices, Stokes' matrices) are constants in the deformation parameters. 
The compatibility of  \eqref{isomonPDE_U}, \eqref{isomonPDE_V} with \eqref{rational_cov_deriv_eq} is equivalent to the set of equations
\begin{subequations}
\bea
\frac{\pa  L}{\pa {t^\nu_{ja} }}  &\& = \frac {\pa U^\nu_{j a } }{\pa z}  + \Big[U^\nu_{ja}  , L\Big] ,\\
\frac{\pa L }{\pa c_\nu}  &\& = \frac {\pa  V^\nu }{\pa z}  + \Big[V^\nu , L \Big],
\eea
\label{ZCE}
\end{subequations}
{\hskip -6 pt} known as {\it zero curvature equations}. (See Section \ref{isomon_deform}.)

 In addition to these, we also have the compatible set of (autonomous)  isospectral deformation equations
\be
\frac{\partial L}{\partial s_a} = \Big[E_{aa}  , L\Big], \quad a=1, \dots , r
\ee
generated  by the exponents  of formal monodromy $\{H^\infty_a\}_{a=1, \dots, r}$  at $\infty$ defined in (\ref{form_mon_infty}), 
giving the $r$-dimensional abelian symmetry group consisting of conjugation by invertible
 diagonal matrices
\bea
\DD_s: L\  &\&\ra \  \tilde{L}:=\DD_s L \DD_s^{-1},
\label{DD_action} 
\\
\DD_s &\&:= \Diag(e^{s_1}, \dots, e^{s_r}) .
\label{DD_def}
\eea
Corresponding to these flows, we may enhance the fundamental system
of  parallel transport equations  (\ref{rational_cov_deriv_eq}),  (\ref{isomonPDE_U}), (\ref{isomonPDE_V})  by
defining 
\be
\tilde{\Psi}(z) := \DD_s \tilde{\Psi}(z),
\label{Psi_tilde_def}
\ee
which satisfies eqs.~(\ref{isomonPDE_U}), (\ref{isomonPDE_V}) and  (\ref{rational_cov_deriv_eq}) with $L$ replaced by $\tilde{L}$,
as well as the further linear equations
\be
\frac{\partial \tilde{\Psi}}{\partial s_a} =E_{aa} \tilde{\Psi}, \quad a=1, \dots, r,
\ee
 implying that the $s_a$ flows  are  both isospectral and isomonodromic.
The parameters $\{s_a\}_{a=1, \dots, r}$, however, are not functions on the phase space,
but genuine independent flow variables.  

Denote the set of isomonodromic deformation parameters  
\be
\Tb:= \Big\{t^\nu_{ja}, c_\nu \Big\}_{\nu = 1, \dots, N, \infty,\, j=1, \dots, d_\nu, \, a=1, \dots, r}.
\label{isotimes}
\ee
As shown in Theorem \ref{zero_curv_Lax}, these consist entirely of Casimir elements for the Lie-Poisson bracket \eqref{split_rational_Rmatrix_PB}.
\begin{remark}
The only further Casimir functions needed to complete the center of the Poisson structure (restricted to rational Lax matrices
(\ref{rational_Lax_matrix}))  are the exponents of formal monodromy $\{t^\nu_{0a}\}_{1\leq \nu \leq N, \, 1\leq a \leq r}$ 
at the finite poles of the rational connection $L(z)d z $ defined in \eqref{form_mon_nu}. 
\end{remark}
It follows from the $R$-matrix theory that the Hamiltonian 
vector fields generated by the spectral invariants (\ref{spectral_Isomon_hamiltonians_t_nu}),  (\ref{spectral_Isomon_hamiltonians_t_inf}) and
(\ref{SchlesHam}) can be interpreted as the commutator terms (\ref{Hamvec_commut}) in eq.~\eqref{ZCE}. 
We will reprove this key fact explicitly  for our Hamiltonians in Section \ref{hamilt_vector fields_comm}, Theorem \ref{zero_curv_Lax}. 
Summarizing, we have the following theorem.    
 \begin{theorem}
\label{Ham__eqs_L}
The Hamiltonian vector fields  corresponding to the Hamiltonians  $H_{t}, \  t\in \Tb$ defined in \eqref{spectral_Isomon_hamiltonians_t_nu}, \eqref{spectral_Isomon_hamiltonians_t_inf}, \eqref{SchlesHam}, and $H^\infty_a$ defined in (\ref{HamVec_inf_a}) are given by the following commutators
\begin{subequations}
\bea
\Xb_{H_{t^\nu_{ja}}}L(z)&\&= \Big\{L(z) , H_{t^\nu_{ja}}\Big\}= \Big[ U^\nu_{ja}(z), L(z)\Big],
\label{X_t_nu_a}
\\
\Xb_{H_{c_\nu}}L(z)&\&=\Big\{L(z) , H_{c_\nu} \Big\}= \Big[V^\nu(z), L(z)\Big],
\label{X_c_nu}
\\
\Xb_{H^\infty_a}L(z) &\&= \{L(z), H^\infty_a\} = [E_{aa}, L],
\label{X_inf_a}
\eea
\label{Ham_vec_fields_commut}
\end{subequations} 
 where the matrices $U^\nu_{ja},V^\nu, E_{aa}$ are expressible as
\begin{subequations}
\bea
U^\nu_{ja}(z)&\&=-(d H_{t^\nu_{ja}})_-, \quad
 V^\nu=- (d H_{c_\nu})_-, \quad \nu=1, \dots, N,  \ j=1, \cdots d_\nu, \ a=1, \dots , r,
 \label{UV_nu_dH}\\
 U^\infty_{ja}(z)  &\&=(d H_{t^\infty_{ja}})_+, \ \, \quad  E_{aa} = (d H^\infty_a)_+,
 \quad j=1, \dots ,  d_\infty,  \ a=1, \dots , r.
  \label{U_infty_dH}
 \eea
 \end{subequations}
\end{theorem} 
We also show (Theorem \ref{d_log_tau_H}) that the isomonodromic $\tau$-function 
defined in \cite{JMU, JM}  (and eqs.~(\ref{tauJMU}), (\ref{isomon_tau})) admits the following representation in terms of spectral invariants 
\be
d \ln \tau_{_{IM}} = \sum_{\nu=1}^N\le( H_{c_\nu} d c_\nu + \sum_{j=1}^{d_\nu} \sum_{a=1}^r H_{t^\nu_{ja}} d t^\nu_{ja} \ri)+ \sum_{j=1}^{d_\infty} \sum_{a=1}^r H_{t^\infty_{ja}} d t^\infty_{ja},
\ee
with the Hamiltonians defined as spectral invariants by formulae\ \eqref{spectral_Isomon_hamiltonians_t_nu}, 
 \eqref{spectral_Isomon_hamiltonians_t_inf},  \eqref{SchlesHam}.

To each of the isomonodromic deformation parameters $t\in  \Tb$ in \eqref{isotimes} 
we associate a vector field  $\nabla_t$   in $ \Gamma(T {\LL_{r, \db}})$ determined by the formulae
\be
\nabla_{t^\nu_{j  a}}L(z) := \frac{\partial U^\nu_{j a}(z)}{\partial z} ,\quad  \quad \nabla_{t^\infty_{j a}} L(z) := \frac{\partial U^\infty_{j a}(z)}{\partial z},\quad
\quad  \nabla_{c_\nu}L(z) := \frac{\partial V^\nu(z)}{\partial z}.
\label{isomon_cond}
\ee
These should be understood as defining the action of $\nabla_t$ on each of the coefficients of $L$ (viewed as linear coordinates on $\LL_{r, \db}$) as well as the position of the poles, extended to arbitrary differentiable functions of $L$ by requiring it to be a derivation.

With these definitions, we can rewrite the isomonodromic equations (\ref{ZCE}) as 
\be
\pa_t L = \nabla_t L+ \Xb_{H_t} L ,\quad  t \in \Tb.
\label{isomon_eqs}
\ee
In Theorems \ref{thmcasi} , \ref{thmflat}  and Proposition \ref{constrank}, the vector fields $\{\nabla_t\}_{t\in \Tb}$ will be shown
to act as coordinate curve directional derivatives, spanning an integrable distribution $\mathfrak T\ss T {\LL_{r, \db}}$ within 
the tangent bundle of ${\LL_{r, \db}}$, generating a maximal regular foliation transversal to the symplectic leaves. In Theorem \ref{thmpoisson} they are
shown to preserve the Poisson brackets, and we interpret them as ``explicit derivatives'' with respect to the various 
isomonodromic times $t= t^\nu_{ja}$ or $c_\nu$.
This is clear, in particular, for the loci $(c_1, \dots, c_N)$ of the poles, where it turns out that $V^\nu$ is precisely 
the (negative of the) singular part of $L(z)$ at $z=c_\nu$ and hence 
\be
\frac{\pa V^\nu(z) }{\pa z} = \frac {\pa^0  L(z)}{\pa c^0_\nu},
\ee
where the derivation $\nabla_{c_\nu}=\frac{\partial^0}{\partial c_\nu^0}$ is the same as ``explicit derivative'' in the obvious sense of differentiating
the rational, matrix valued function $L(z)$ with respect to its pole locations while keeping all other parameters fixed.

For the higher Birkhoff invariants $\{t^\nu_{ja}\}_{\nu = 1, \dots, N, \, j=1, \dots, d_\nu, \, a=1, \dots, r}$,  
it is not at all obvious in which sense eq.~(\ref{isomon_cond}) may be thought of
as defining ``explicit derivatives''. The key point is that, in order to interpret the $\nabla_t$'s  as such,
we must {\em prove} both that they mutually  {\it commute} and that, when applied to the invariants $\{t^\nu_{ja}, c_\nu\}$, they really
act as directional derivatives along flow lines in which all the remaining transversal coordinates are kept fixed.
 These results are included in the following theorem, which combines those of Theorems \ref{thmcasi}, \ref{thmflat}, 
 and  \ref{thmpoisson} of Section  \ref{birkhoff_conn} and Proposition \ref{constrank} of Section \ref{Further}.
Combined with Theorem \ref{Ham__eqs_L}, this shows how the Hamiltonian vectors fields of eqs.~(\ref{Ham_vec_fields_commut}),
when added to the ``explicit'' derivative  vector fields,  produce the zero curvature equations (\ref{ZCE})  that guarantee consistency of the equations (\ref{rational_cov_deriv_eq}), (\ref{isomonPDE_U}), (\ref{isomonPDE_V}),  and ensure the invariance of the (generalized) monodromy 
of the system (\ref{rational_cov_deriv_eq})  under changes in the parameters $\{t^\nu_{ja},  c_\nu\}$.
  \begin{theorem}
\label{rational_isomon_systems}
\begin{enumerate}
\item If $s, t$ denote any two isomonodromic times $s,t \in \Tb $ then 
\be
\nabla_{s} t=\delta_{st}.
\ee 
\item The vector fields $\{\nabla_t\}_{t \in \Tb}$ in \eqref{isomon_cond} commute amongst themselves and  span an integrable distribution of constant rank in the space of rational Lax matrices of the form \eqref{rational_Lax_matrix}. 
\item The  $R$-matrix Poisson structure is invariant with respect to the local flows generated by  the $\nabla_t$'s.
That is,
\be
\nabla_t \{f, g\} = \{\nabla_t f, g\} + \{f, \nabla_t\, g \} \quad \forall\,  t\in \Tb, \ f, g \in C^\infty(\LL_{r, \db}).
\ee
\end{enumerate} 
\end{theorem} 

\begin{remark} A generalization of the ``explicit'' derivative vector fields $\nabla_t$  for isomonodromic systems
 over an arbitrary Riemann surface is defined in {\rm \cite{Hu}}, without further developing their properties. \end{remark} 
%%%%%%%%%%%%%%%  Section 2. Hamiltonian theory of $P_{II}$  %%%

\section{An illustrative example: Hamiltonian theory of $P_{II}$ }
\label{P_II_hamiltonian}

Before proceeding to the general  case, we recall the example of  the second Painlev\'e  transcendent $P_{II}$
as the simplest illustration of an isomonodromic deformation system with $r=2$ and  a polynomial Lax matrix of degree $2$. 

The $P_{II}$ equation is:
\be
u'' = 2 u^3 + t u + \alpha,
\label{P_II}
\ee
with arbitrary constant $\alpha$. 
To interpret (\ref{P_II}) as an isomonodromic deformation equation \cite{JM, HR}, we introduce a pair of $2\times 2$ matrices $(L(z)$, 
$U(z))$, which are second and first degree  polynomials in $z$, parametrized as
\bea
L(z) &\&:= z^2 
\begin{pmatrix} 1 & 0 \cr 0 & -1 \end{pmatrix}
 +z
 \begin{pmatrix} 0 & - 2y_1 \cr x_2 & 0 \end{pmatrix}
+ \begin{pmatrix} x_2 y_1 + {t\over 2} & - 2 y_2 \cr x_1 & -x_2y_1- {t\over 2} \end{pmatrix}
\\
U(z)&\&:=
 {z\over 2} 
 \begin{pmatrix} 1 & 0 \cr 0 & -1 \end{pmatrix} 
 + {1\over 2} 
 \begin{pmatrix} 0 & -2y_1 \cr x_2 &  0 \end{pmatrix}.
\eea
The overdetermined pair of matrix equations
\bea
{\partial \Psi(z)\over \partial z} &\&= L(z)\Psi(z)  , \\
{\partial \Psi(z) \over \partial t} &\&= U(z)\Psi(z) 
\label{P_II_def_eq}
\eea
is compatible if and only if $L(z)$ satisfies the evolution equation
\be
{\partial L \over \partial t  } =  [U(z), L(z)] + {1\over 2} \begin{pmatrix} 1 & 0 \\ 0 & -1 \end{pmatrix}
  =[U(z), L(z)] + {\partial U(z)\over\partial z},
\label{PII_ZCE}
\ee
which  guarantees  the invariance of the (generalized) monodromy of the meromorphic covariant derivative operator
\be
{\partial \over \partial z} - L(z)
\ee
 under the  $t$-deformations generated by (\ref{P_II_def_eq}).

Eq.~(\ref{PII_ZCE}) is of Hamiltonian type \cite{JM, Ok1, Ok2}
with respect to the canonical $1$-form 
 \be
 \theta := y_1 d x_1 + y_2 d x_2.
 \label{canon_1_form_2}
 \ee
with $t$ viewed as a Casimir element.
 The map
 \be
 (t, x_1, x_2, y_1, y_2) \ra L(z)
 \ee
 is then a Poisson map with respect to the Poisson bracket \eqref{split_rational_Rmatrix_PB} on the dual $L^*\grgl_R(2)$ of the loop algebra
 $L\grgl_R(2)$  
 defined by the rational split $R$-matrix.
 The nonautonomous Hamiltonian $H_{II}$ is the spectral invariant
\bea
H_{II} &\&= \frac{1}{4}  \res_{z=0} z^{-1}\tr(L^2(z)) -\frac{t^2}{8} =\frac{1}{ 2} \left( x_2^2y_1^2 + t x_2 y_1 - 2 x_1 y_2\right) ,
\label{H_II_def}
\eea
and we have 
\be
- (d H_{II})_- = U(z),
\ee 
as required by the $R$-matrix theory. Taking into account
the explicit dependence of $L(z)$ on the parameter $t$, viewed as a function on the 
phase space, note that $t$ is in fact a spectral invariant Casimir function,  which can be computed as 
\be
 t= \frac{1}{2}  \res_{z=0} z^{-3}\tr\left(L^2(z)\right)d z  = 2t^\infty_{11},
 \label{t_res_eq}
\ee
and hence identified with the first Birkhoff invariant.
That the zero curvature equations (\ref{PII_ZCE}) are of the deformed Hamiltonian form (\ref{isomon_eqs}) follows 
from the fact that the explicit $t$ derivative of $L(z)$ is
\be
\frac{\partial^0 L}{\partial t^0}=\nabla_t(L) = {1\over 2} \begin{pmatrix} 1 & 0 \\ 0 & -1 \end{pmatrix} = \frac{\partial U}{\partial z}.
\ee
To check that formulae (\ref{t_res_eq} ), (\ref{H_II_def}) coincide with the definitions  (\ref{princ_part_coeffs_inf}), 
(\ref{spectral_Isomon_hamiltonians_t_inf})  of  $t$  as twice the Birkhoff invariant $t^\infty_{11}$ 
and  $H_{II}$ as the ``dual'' Hamiltonian $H_{t^\infty_{11}}$, 
note that the meromorphic differential $\lambda(z) dz$  is defined, on its two branches, by the formulae
\be
\lambda = \pm \sqrt {-\det(L)} = \pm\le (z^2 + \frac t 2 - \frac {x_1y_1 + x_2 y_2}z + \frac {H_{II}}{z^2} + \dots\ri).
\label{lambda_exp_P_II}
\ee

Choosing new canonical coordinates
\be
u :=\frac{x_1}{ x_2},\quad v:= x_2 y_1,   \quad w:= \ln x_2,  \quad a:=x_1y_1+ x_2y_2,
\ee
the canonical $1$-form (\ref{canon_1_form_2}) becomes
\be
\theta= v d u+ a d w,
\ee
and the Hamiltonian is
\be
H _{II}= \frac{1}{2 } v^2 + {1\over2} (t +2 u^2) v- a u .
\ee
Note that $a$ is an autonomous spectral invariant of the Lax matrix, and hence a conserved quantity.
It can be written equivalently as
\be
a = -\frac{1}{4}\res_{z=0} z^{-2} \tr(L(z))^2 = H^\infty_1 = - H^\infty_2,
\ee
which   is equal to the exponent of formal monodromy $t^\infty_{01}$ at $\infty$.
The Hamiltonian flow it generates is the  group of scaling symmetries
\bea
f_s: (x_1, y_1,  x_2, y_2, t)&\& \ra \{e^s x_1, e^{-s} y_1,  e^s x_2, e^{-s} y_2, t\} , 
\label{scaling_P_II}
\\
f_s: L(z) &\&\ra e^{s\sigma_3}L(z) e^{-s\sigma_3}.
\label{scaling_Lax_P_II}
\eea

The corresponding canonically conjugate position variable $w$ is thus an ignorable coordinate, 
 and Hamilton's equations reduce to the two-dimensional system
\be
\frac{d u}{d t  } = v + u^2+{t\over 2}, \quad {d v \over d t  } = -2u v +a.
\label{P_II_first_order}
\ee
Eliminating the $v$ variable to obtain an equivalent second degree equation for $u$ gives the 
$P_{II}$ equation (\ref{P_II}) with $\alpha = a - {1\over 2}$.
The associated isomonodromic $\tau$-function is  related to the Hamiltonian $H_{II}$ by
\be
\d(\ln \tau) = H_{II} \, d t  .
\label{tau_P_II_H}
\ee

%%%%%%%%%%%%%%%  Section 3. Isomonodromic deformations for rational Lax matrices  %%%

\section{Isomonodromic deformations: general rational Lax matrices}
\label{secMonodromy}

%%%%%%%%%%%%%%%  Section 3.1.  Rational Lax matrices %%%

\subsection{Rational Lax matrices}
\label{rational_Lax_mat}

We start our analysis of the general case by defining notation and reviewing the underlying notions. 
Consider the space of matrices of the form \eqref{rational_Lax_matrix}:
\be
\label{Lhat}
{\LL_{r, \db}}:= \bigg\{&
L(z) =
 -\sum_{j=0}^{d_\infty-1} L^{\infty}_{j+2} z^j + \sum_{\nu=1}^N
\sum_{j=1}^{d_\nu+1} \frac {L^{\nu}_{j}}{(z-c_\nu)^j},  \cr 
& L_{d_\infty+1}^{\infty}\in \mathfrak h_{reg} \ss \grgl(r),  \  L_{d_\nu+1}^{\nu}\in {\frak g}_{reg} \ss \grgl(r), \ \ c_\nu\neq c_\mu, \ \nu\neq \mu\bigg\}
\ee

Here $\frak h_{reg}$ denotes the set of ad-regular {\it diagonal} matrices (i.e.~with distinct eigenvalues) in $\grgl(r)$, 
while $\frak g_{reg}$ is the set of all $\grgl(r)$  matrices with distinct eigenvalues. The rationale behind the indexing of the matrices
$\{L^\nu_j, L^\infty_{j+2}\} $ is  that  the subscript should coincide with the order of the pole of the  matrix-valued differential form $L(z) d z$ 
near the corresponding pole. In particular, if we define  local coordinates near its poles by 
\be
\zeta_\nu = \le\{
(z-c_\nu), \ \ \nu=1,\dots, N
\atop
\frac 1 z ,\ \ \ \nu=\infty,
\ri.
\label{zetanu}
\ee
 the singular part of $L(z)  d z$  near each pole  can be written in a uniform manner as 
\be
\left(L(z) \right)_{sing}d z = \sum_{j=1}^{d_\nu+1}\frac{ L^\nu_{j}}{\zeta_\nu^j}d \zeta_\nu, \quad \nu=1,\dots,N,\infty.
\ee

We now recall  the classical description  \cite{BJL,  JMU, JM, Wa, Sib} of formal solutions near an isolated
 singular point of a linear first order system of ordinary differential equations  of the form (\ref{rational_cov_deriv_eq}).

\begin{proposition}
\label{propODE}
Consider the linear system of first order ODE's in the complex plane
\be 
\frac{d\Psi(z)}{dz} = L(z) \Psi(z),\quad L(z) \in {\LL_{r, \db}}, \quad \Psi(z) \in \grGl(r),
\label{ODE_L_z}
\ee
with ${\LL_{r, \db}}$ defined in \eqref{Lhat}, and  $L^\infty_{d_\infty +1}$ chosen to be diagonal. 
In terms of the local parameters $\zeta_\nu$ defined in \eqref{zetanu}, 
there exist local formal series solutions of the form 
\be
\Psi^\nu_{\text{\tiny form}}(z) =  Y^\nu (\zeta_\nu) {\rm e}^{T^\nu(\zeta_\nu)} , \ \ \ Y^\nu(\zeta_\nu) := G^\nu\le(\Ib + \sum_{j\geq 1} Y_j^\nu {\zeta_\nu}^j\ri),
\label{formal_Psi}
\ee
in a punctured neighbourhood of each of the singular points $\{z=c_\nu\}$  (including $\infty$),
where  $T^\nu(\zeta_\nu)\in \mathfrak h_{reg}$ is a  diagonal matrix of the form 
\be
T^\nu (\zeta_\nu) = \sum_{j=1}^{d_\nu }\frac { T_j^\nu}{j {\zeta_\nu}^j} + T_0^\nu \ln \zeta_\nu, \quad  \ T^\nu_{d_\nu } = -(G^\nu)^{-1}L^\nu_{d_\nu+1} G^\nu,
\label{formal_T_nu}
\ee
for $\nu=1,\dots, N,\infty$. 
The columns of the invertible matrices $G^\nu\in GL(r,\C)$ are the independent eigenvectors of $L^\nu_{d_\nu+1}$ and $G^\infty=\Ib$.
\end{proposition}

 We use the following notation for the diagonal values
\be
T^{\nu}_{j} = {\rm diag}(t^{\nu}_{j 1},\dots,t^{\nu}_{j  r}), \quad j=0, \dots, d_\nu ,
\label{T_mu_j}
\ee
so 
\be
T^{\nu}(\zeta_\nu) = \sum_{j=1}^{d_\nu} \sum_{a=1}^r t^{\nu}_{j a} E_{aa} \frac1{j\zeta_\nu^{j}} + \sum_{a=1}^r t^{\nu}_{0a}E_{aa} \ln \zeta_\nu,
\label{T_nu}
\ee
where $E_{ab}$ is the elementary matrix whose only nonzero entry is $1$ in the $ab$ position
and
\be
t^{\nu}_{j a} \neq t^{\nu}_{j b} \ \  \text{for } a\neq b, \ \ j=1, \dots, d_\nu, \ \ \nu =1, \dots, N, \infty.
\ee 
\begin{definition}
\label{birkhoff_invar}
 The  entries  $(t^{\nu}_{j 1},\dots,t^{\nu}_{j  r})$ of the diagonal matrices $\{T^\nu_j\}_{j=0, \dots, d_\nu}$  
are called the  {\rm Birkhoff invariants}. In particular, the entries $(t^{\nu}_{0 1},\dots, t^{\nu}_{0  r})$ in the matrices $T^\nu_0$
are called the {\rm exponents of formal monodromy} \cite{BJL, JMU} and the entries $(t^{\nu}_{ja},\dots,t ^{\nu}_{j r})$  for $1\leq j \leq d_\nu$
the {\rm higher Birkhoff invariants}. The integers $\db=(d_1, \dots d_N, d_\infty)$ are the {\em Poincar\'e ranks} of the various singular 
points $\{c_\nu\}_{\nu=1, \dots, N, \infty}$.

\end{definition}
 
 %%%%%%%%%% 3.2. Isomonodromic deformations %%%%
\subsection{Isomonodromic deformations and zero curvature equations}
\label{isomon_deform}

The higher {\rm Birkhoff invariants} $\{t^\nu_{ja}\}_{\nu=1, \dots, N, \infty, \,j=1, \dots d_\nu,\, a=1, \dots r}$,   
together with the  loci $\{c_\nu\}_{\nu =1, \dots, N}$  of the finite poles  will serve as  isomonodromic deformation parameters.
To each, there corresponds an infinitesimal deformation equation which we now recall,  following \cite{JMU, JM}.
For $ \nu=1\dots, N,\infty $ and  $j=1,\dots, d_\nu$,  define the following matrices in terms of the Birkhoff 
invariants in $T^\nu(\zeta_\nu)$ and the formal series $Y^\nu(\zeta_\nu)$ in \eqref{formal_Psi} near $z=\zeta_\nu$. 
\bea
U^\nu_{ja}(z; L)&\&:= \le(Y^\nu(\zeta_\nu) \frac {\pa  T^\nu(\zeta_\nu)}{\pa t^\nu_{ja}} (Y^\nu(\zeta_\nu))^{-1}\ri)_{sing}
\hspace{-10pt} = \le(Y^\nu(\zeta_\nu) \frac { E_{aa} }{j \zeta_\nu^j } (Y^\nu(\zeta_\nu))^{-1}\ri)_{sing},
\label{defU}
 \\
V^\nu(z;L)&\& :=  \le( Y^\nu(\zeta_\nu) \frac {\pa  T^\nu(\zeta_\nu)}{\pa c_\nu } (Y^\nu(\zeta_\nu))^{-1}\ri)_{sing}\hspace{-10pt}
=- \le( Y^\nu(\zeta_\nu) \frac {d T^{\nu}(\zeta_\nu)}{d z} (Y^\nu(\zeta_\nu))^{-1}\ri)_{sing}= -\sum_{j=1}^{d_\nu+1}
\frac {L^\nu_{j}} {(z-c_\nu)^j}.
\label{defV} \cr
&\&
\eea
where  $(\cdot )_{sing}$ denotes the principal part\footnote{The notation $(\cdot)_{sing}$ means the principal part 
 at a particular point $c_\nu\in \Pbb^1$, which should be clear from the context or, if not,  will be specified.}
  of the Laurent series $(\cdot )$ near $z=c_\nu$ .
For a Laurent series 
\be
f(z) = \sum_{n\in\Zbb} a_n (z-c)^n
\ee
centered at $c\in \Cbb^1$  the principal part  may be  expressed as
\be
\label{sing}
\big(f(z)\big)_{sing} :=  \sum_{n\ge 1} \frac{a_{-n}}{ (z-c)^n} =\res_{w=c} \frac{f(w)d w}{z-w}.
\ee
At $c_\infty=\infty$,  the principal part is simply the polynomial part of the expression  (including the constant term). 
The main result of \cite{JMU, JM} is that if the following system of equations is satisfied,
\begin{subequations}
\bea
\frac {d \Psi(z)}{d z}  &\& = L(z) \Psi(z),
\label{L_z_eq} \\
\ \frac {\pa  \Psi(z)}{\pa t^\nu_{ja}} &\&= U^\nu_{ja}(z) \Psi(z) , \quad j=1, \dots, d_\nu ,
\label{t_nu_j_a_def_eq}\\
\frac {\pa  \Psi(z) }{\pa c_\nu}&\&= V^\nu(z) \Psi(z), \quad \nu =1, \dots , N
\label{c_nu_def_eq}\
\eea
\end{subequations}
(where, for brevity, we write  $U^\nu_{ja}(z)$, $V^\nu(z)$ for   $U^\nu_{ja}(z; L),$ $V^\nu(z,;L)$),
the  {\it generalized monodromy} (including the values of the Stokes matrices defined in a neighbourhood of each irregular singular point) is independent of
the deformation parameters $\{t^\nu_{ja}, c_\nu\}$. 

This means that the matrices $\{L, U^\nu_{ja},V^\nu\}$, restricted to the isomonodromic solution
manifold, satisfy the {\it zero curvature equations} (\ref{ZCE}). Equivalently, defining the connection form
\be
\Omega(z; {\bs t}, \bs c)  :=-L(z)d z -  \sum_{\nu=1}^{N,\infty} \left(\sum_{j=1}^{d_\nu}\sum_{a=1}^r U^\nu_{ja} d t^\nu_{ja} +  V^\nu d c_\nu \right) ,
\label{Omegadef}
\ee
the corresponding curvature vanishes
\be
\delta \Omega+ [\Omega,\Omega] =0,
\label{zero_curv_eq}
\ee
where
\be
\delta  :=  d z \frac {\pa}{\pa z} + \sum_{\nu=1}^N d c_\nu \frac{\partial}{\partial c_\nu} +   \sum_{\nu=1}^N  \sum_{j=1}^{d_\nu}\sum_{a=1}^r d t^{\nu}_{j a} \frac{\partial}{\partial t^{\nu}_{j a}}  +  \sum_{j=1}^{d_\infty}\sum_{a=1}^r d t^{\infty}_{j a} \frac{\partial}{\partial t^{\infty}_{j a}}
\label{delta_differential}
\ee
is the total differential operator acting on scalar functions of $\{z, t^\nu_{ja}, c_\nu\}$ 
extended, as usual, to the exterior differential on the space of differential forms.

We denote  differentials with respect to the parameters at each pole location as
\bea
{\rm d}_\nu := d c_\nu \frac \pa{\pa c_\nu}+ \sum_{j=1}^{d_\nu} \sum_{a=1}^r 
d t  ^{\nu}_{ja} \frac \pa{\pa t^{\nu}_{ja}}, \quad
{\rm d}_\infty := \sum_{j=1}^{d_\infty} \sum_{a=1}^r  d t  ^{\infty}_{ja} \frac
\pa{\pa t^{\infty}_{j a}}\ .
\label{d_nu_differentials}
\eea
so
\be
{\rm d} =  \sum_{\nu=1}^N {\rm d}_\nu  + {\rm d}_\infty
\ee
is the total differential on the space of deformation parameters.
It is proved in  \cite{JMU, JM} that the differential
\be
\omega_{_{IM}} :=-\sum_{\nu=1, \dots, N, \infty} \res_{z=c_\nu}
 \Bigg(\tr \le( 
(Y^{\nu}(\zeta_\nu))^{-1} \pa_z Y^{\nu}(\zeta_\nu) {\rm d}_\nu T^{\nu}(\zeta_\nu) \ri)d z\Bigg) 
\label{tauJMU}
\ee
is  closed when restricted to the solution manifold of the isomonodromic equations and hence locally exact. 
The isomonodromic $\tau$-function $\tau_{IM}$ is locally defined, up to a parameter independent normalization, by 
\be
{\rm d} \ln \tau_{_{IM}} := \sum_{\nu=1, \dots, N, \infty}{\rm d_\nu}  \ln \tau_{_{IM}} = \omega_{_{IM}}.
\label{isomon_tau}
\ee
Note that this  actually defines a section of a line bundle over the solution space of the isomonodromic equations,
 and that the ``function''  $\tau_{_{IM}}$ is not, in general, single valued.
 Moreover, it is only defined up to a multiplicative factor that does not depend on the deformation parameters $\{t^\nu_{ja}, c_\nu\}$.

%%%%%% Subsection 3.3 Expression in terms of spectral invariant %%%%

\subsection{Expression in terms of spectral invariants}
\label{spectralinvars}

Our first result consists of a representation of formula (\ref{isomon_tau})  determining the isomonodromic $\tau$-function in 
terms of {\it spectral invariants}.  This was announced previously \cite{H4} in the form expressed in Theorems \ref{lemmaeigenvectors},
 \ref{d_log_tau_H} below, appeared in slightly different contexts in \cite{BertoMo, Hu} and more recently was re-derived in \cite{Yam}.
 The proofs given here are complete and self-contained.

We define the spectral curve  $\CC$ as the compactification of the affine curve $\CC_0$ cut out by the characteristic equation 
(\ref{char_eq_L}) of $L(z)$ obtained by adding $r$ points over each finite $z=c_\nu$, and over $z=\infty$.
This allows us to view $\lambda\, dz$ as a globally defined meromorphic differential on the compactified spectral curve $\CC$,
with the locally defined Laurent series for the functions $\lambda_1(z), \dots , \lambda_N(z)$ in punctured neighbourhoods of the
points $\{z=c_\mu\}_{\nu =1 , \dots, N, \infty}$  representing the values of $\lambda$ near the $r$ distinct points of $\CC$ projecting to the punctured
neighborhood of $z=c_\nu$.
By our requirement that the highest degree terms  $L^\nu_{d_\nu+1}$'s all  have {\it distinct} eigenvalues, 
the eigenvalues $\{\lambda^\nu_a(\zeta_\nu)\}_{a=1, \dots r}$ of $L(z)$,
 in a neighbourhood of each of the poles $\{z=c_\nu\}_{\nu =1, \dots, N, \infty }$, will be expressed by $r$ distinct Laurent series expansions,
with poles of order $d_\nu+1$ for the poles at the finite points $\{z=c_\nu\}_{\nu =1, \dots, N}$ and $d_\infty-1$ for the pole at infinity. 
(As indicated above, the difference in indexing at finite poles and at $\infty$ is due to the fact that the objects of interest 
are the matrix valued $1$-form $L(z)\,d z$ and the meromorphic differential $\lambda \, d z$,  which have poles of order $d_\nu+1$
at  the $z=c_\nu$'s, including $c_\infty=\infty$, since $d z$ has a double pole at $\infty$.)

The first step is to prove  formulae \eqref{princ_part_coeffs_nu}, \eqref{princ_part_coeffs_inf},  \eqref{spectral_Isomon_hamiltonians_t_nu}, \eqref{spectral_Isomon_hamiltonians_t_inf} and \eqref{SchlesHam}; i.e.,  to express the Birkhoff invariants $\{t^\nu_{ja}, c_\nu\}$ and their ``dual'' partners, 
the Hamiltonians $\{H_{t^\nu_{ja}}, H_{c_\nu}\}$, in terms of the spectral curve $\CC$ defined in  \eqref{char_eq_L}.
In order to treat the local behaviour near the finite poles $\{z=c_\nu\}_{\nu =1, \dots, N}$ and at $z= c_\infty$ uniformly in what follows,
 we will always consider the matrix-valued $1$-form $L(z) \,d z$, with $\zeta_\nu$  chosen 
 as one of the local coordinates (\ref{zetanu}).  Note that
\be
\frac{d z}{d \zeta_\nu} = 1, \quad \nu=1, \dots, N, \quad \frac{d z}{d \zeta_\infty} = - \frac 1 {\zeta_\infty^2}.
\label{dz_dzeta_nu}
\ee
 
 Choose a suitably normalized invertible matrix $P(z)$, whose columns are the linearly
  independent eigenvectors of $L(z)$, and which has local Taylor expansions near to each $z=c_\nu$
\be
P(z)|_{\text{near } z=c_\nu} =:P^\nu(\zeta_\nu) = G^\nu \Big(\Ib + F^\nu_1 \zeta_\nu + F^\nu _2\zeta_\nu ^2 + \dots \Big),
\label{P_nu_exp}
\ee
where $G^\nu$ is the same invertible matrix as in \eqref{formal_Psi}.
These satisfy 
\begin{subequations} 
\bea
L(z) P^\nu(\zeta_\nu) &\&= P^\nu (\zeta_\nu) \Lambda^\nu(\zeta_\nu), \quad \nu=1, \dots, N,
\label{LP_Lambda_nu}  \\
  L(z) P^\infty(\zeta_\infty) &\& = -\zeta_\infty^2 P^\infty (\zeta_\infty) \Lambda^\nu(\zeta_\infty)  
\label{LP_Lambda_inf}
\eea
\end{subequations} 
 in  punctured neighbourhoods of each pole $\{z=c_\nu\}_{\nu =1, \dots, N,  \infty}$, where
 \begin{subequations}
\bea
\Lambda^\nu(\zeta_\nu) &\&= \diag(\lambda_1(z), \dots, \lambda_r(z)) =  \diag(\lambda^\nu_1(\zeta_\nu), \dots, \lambda^\nu_r(\zeta_\nu)), 
 \quad \nu=1, \dots, N,
 \label{Lambda_def_nu} \\
\Lambda^\nu(\zeta_\infty) &\&=-z^2 \diag(\lambda_1(z), \dots, \lambda_r(z)) = -\zeta_\infty^{-2}\,  \diag(\lambda^\infty_1(\zeta_\infty), \dots, \lambda^\infty_r(\zeta_\infty))  
\label{Lambda_def_inf}
\eea
\end{subequations}
is the diagonal matrix of eigenvalues, which are the $r$ (distinct)  solutions of the characteristic equation (\ref{char_eq_L}), as local Laurent series
near each pole $z=c_\nu$.
We then have the following key result, which implies formulae\ \eqref{princ_part_coeffs_nu}, \eqref{princ_part_coeffs_inf}
 
\begin{theorem}(Residue formulae for the Birkhoff invariants.)
\label{lemmaeigenvectors}
\begin{enumerate}
\item For a suitable choice of normalization of the eigenvectors forming the columns of $P^\nu(\zeta_\nu)$,  
the  matrix $Y^\nu(\zeta_\nu)$, with formal expansion \eqref{formal_Psi},  coincides with the analytic series (\ref{P_nu_exp}) 
for $P^\nu(\zeta_\nu)$ up to terms of order $\mathcal O(\zeta_\nu^{d_\nu+1} )$.
\item The matrix $ \frac {d  T^\nu(\zeta_\nu)}{d \zeta_\nu}$ equals the principal part of the local Laurent
series of the matrix $\Lambda^\nu(\zeta_\nu)$ of eigenvalues near $z=c_\nu$
\be
 \frac {d T^\nu}{d \zeta_\nu}(\zeta_\nu) =  \big(\Lambda^\nu(\zeta_\nu)\big)_{sing},
 \label{Lambda_sing_T_prime}
\ee
\end{enumerate}
where 
\begin{subequations}
\bea
\lambda^\nu_a(\zeta_\nu) &\&=  -\sum_{j=0}^{d_\nu}\frac{ t^\nu_{ja}}{\zeta_\nu^{j+1}} +   \OO(1),  \quad \nu=1, \dots, N,  
\label{lamba_nu_a}
\\
\lambda^\infty_a(\zeta_\infty) &\&=  \sum_{j=0}^{d_\infty}\frac{ t^\infty_{ja}}{\zeta_\infty^{j-1}} +   \OO(\zeta^2_\infty),
\eea
\label{lamba_exp_a}
\end{subequations}
and hence 
\begin{subequations}
\bea
t^\nu_{j  a} &\&=- \res_{z=c_\nu } (z-c_\nu)^j \lambda_a(z) d z , \quad \nu = 0, \dots, N,  \quad j=1, \dots, d_\nu,
\label{t_nu_ja_res}
 \\
  t^\infty_{j  a} &\& = -\res_{z=\infty } z^{-j} \lambda_a(z) d z , \quad j=0, \dots d_\infty.
  \label{t_inf_ja_res}
\eea
\label{spectral_res_principal}
\end{subequations}
\end{theorem} 
The results of Theorem \ref{lemmaeigenvectors} should be compared with Remark 1 and eqs.~(2.31)-(2.32) 
 in  \cite{JMU} and  eq. (4.5) in \cite{IFK}.
\begin{proof}
Near each pole $z=c_\nu$, the locally analytic matrix $P^\nu(\zeta_\nu)$  of eigenvectors of $L(z)$  is  given by the series (\ref{P_nu_exp}).
In a punctured neighbourhood of the pole  $z=c_\nu$,  where the formal series expansions  \eqref{formal_Psi}, \eqref{formal_T_nu} hold,
we can express  (\ref{ODE_L_z}) as 
\be
\label{formal2}
 \frac{d Y^\nu}{d \zeta_\nu}   + Y^\nu 
\frac {d T^\nu }{d \zeta_\nu}
 = L(z) \frac {d z}{d \zeta_\nu} Y^\nu.
\ee

Expanding $L(z)\frac{ d z}{d \zeta_\nu}  $ in the local coordinate $\zeta_\nu $ (where $\frac{dz}{d\zeta_\nu}$ is given by (\ref{dz_dzeta_nu})),
we have
\be
L(z) \frac {d z}{d \zeta_\nu}= \sum_{j=1}^{d_\nu+1}  \frac {L^\nu_{j}} {\zeta_\nu^j} + \mathcal O(1).
\ee
Near each $z=c_\nu$, the eigenvector equation can be written in matrix form as (\ref{LP_Lambda_nu}), (\ref{LP_Lambda_inf}),
where
\be
\Lambda^\nu(\zeta_\nu)= \diag(\lambda^\nu_1(\zeta_\nu), \dots, \lambda^\nu_r(\zeta_\nu))
\ee
 is  the diagonal matrix with entries the distinct solutions of the characteristic equation (\ref{char_eq_L}),  whose local Laurent series' have a pole of order $d_\nu+1$ at $\zeta_\nu=0$.   Any (formally analytic) series $P^\nu(\zeta_\nu)$ that satisfies \eqref{LP_Lambda_nu},  \eqref{LP_Lambda_inf} with  diagonal $\Lambda^\nu(\zeta_\nu)$ or $\Lambda^\infty(\zeta_\infty)$ is an eigenvector matrix. Note that the singular part of \eqref{formal2} is identical to the singular part of \eqref{LP_Lambda_nu} or
\eqref{LP_Lambda_inf} and contains the expansion up to order $d_\nu$. Thus the first $d_\nu$ terms in the expansion 
of $Y^\nu(\zeta_\nu)$ coincide with the eigenvector matrix. 
Eqs.~(\ref{LP_Lambda_nu}), (\ref{LP_Lambda_inf}) then imply eq.~(\ref{Lambda_sing_T_prime}), concluding the proof.
\end{proof}

As a consequence of Theorem \ref{lemmaeigenvectors}, it also follows that formulae  \eqref{spectral_Isomon_hamiltonians_t_nu},
\eqref{spectral_Isomon_hamiltonians_t_inf}, \eqref{SchlesHam} are equivalent  to the coefficients 
in  \eqref{tauJMU}, \eqref{isomon_tau},  expressed as spectral invariant residues, 
 viewed as Hamiltonian functions on the phase space $\LL_{r, \db}$, evaluated on the isomonodromic solution manifold.
 
\begin{theorem}
\label{d_log_tau_H}
The components of formulae \  \eqref{tauJMU}, \eqref{isomon_tau}
 \begin{subequations}
\bea
\pa_{t^\nu_{ja}} \ln \tau_{IM} &\&=- \res_{z=c_\nu} \tr \le(( Y^\nu)^{-1} \frac {d  Y^\nu}{d z}  \frac {\pa T^\nu}{\pa t^\nu_{j a} }  \ri)d z
\label{tnuja} \\
 \nu  =1,\dots, N, \infty, &\& \quad j=1,\dots, d_\nu, \quad a= 1,\dots, r, \cr
 &\& \cr
\pa_{c_\nu} \ln \tau_{IM} &\& =- \res_{z=c_\nu} \tr \le(( Y^\nu)^{-1} \frac {d Y^\nu }{d z}  \frac {\pa T^\nu }{\pa c_\nu } \ri)d z \cr
 \nu &\& =1,\dots, N, \
\label{cnu}
\eea
\end{subequations}
are equivalent to the following expressions in terms of residues at the singular points  of the meromorphic differential $\lambda\,  d z$
\begin{subequations}
 \bea
\pa_{t^\nu_{ja}} \ln \tau_{IM} &\&= -\frac 1{j} \res_{z=c_\nu} \frac{1}{ \zeta_\nu^{j}} \lambda_a(z)d z = H_{t^\nu_{j a}},
 \label{H_t_nu_ja} \\
&\& \cr
 \nu=1,\dots, N, \infty, &\& \quad j=1,\dots, d_\nu, \quad a= 1,\dots, r,
 \nonumber \\
 &\& \cr
\pa_{c_\nu} \ln \tau_{IM} &\&=\frac 1 2 \res_{z=c_\nu}\tr \Big(L^2(z)\Big) d z = H_{c_\nu},
\label{H_c_nu} \\
 \nu &\& =1,\dots, N, 
 \nonumber
\eea
\end{subequations}
where the second equalities in (\ref{H_t_nu_ja})  (\ref{H_c_nu}) are the definitions (\ref{spectral_Isomon_hamiltonians_t_nu}),
 (\ref{spectral_Isomon_hamiltonians_t_inf}), (\ref{SchlesHam}) of the  Hamiltonians $\{H_{t^\nu_{j a}}, H_{c^\nu}\}$ as spectral invariants,
 and hence
 \begin{subequations}
 \bea
\lambda^\nu_a(\zeta_\nu) &\&= - \sum_{j=1}^{d_\nu}\frac{ t^\nu_{ja}}{\zeta_\nu^{j+1}}  -  \frac{ t^\nu_{0a}}{\zeta_\nu}
-  \sum_{j=1}^{d_\nu} jH_{t^\nu_{ja}} \zeta_\nu^{j-1}
+ \OO(\zeta_\nu^{d_\nu}),   \quad \nu=1, \dots, N, 
\label{lambda_nu_a_exp}  \\
\lambda^\infty_a(\zeta_\infty) &\&=  \sum_{j=1}^{d_\nu}\frac{ t^\infty_{ja}}{\zeta_\infty^{j-1}} 
+ t^\infty_{0a}\zeta_\infty  +\sum_{j=1}^{d_\infty} j H_{t^\infty_{ja}} \zeta_\infty^{j+1}
+ \OO(\zeta_\infty^{d_\infty+2}).
\label{lambda_inf_a_exp}   
\eea
\end{subequations}
\end{theorem} 
The results of Theorem \ref{d_log_tau_H} should be compared with Remark 5.2 and eqs.~(5.1), (5.1)' in \cite{JMU} 
and Lemma 4.1 in  \cite{IFK}.
\noindent
\begin{proof}
  In terms of the  Laurent expansion in the local parameter $\zeta_\nu$ near to $z=c_\nu$,  
eq.~\eqref{ODE_L_z} implies 
\be
(Y^\nu)^{-1} \frac {d Y^\nu}{ d\zeta_\nu} =  \frac {d T^\nu } {d \zeta_\nu} +  \frac {d z}{d \zeta_\nu}(Y^\nu)^{-1} L Y^\nu.
\ee
Substituting this expression into the RHS of \eqref{tnuja}, we obtain 
\bea
- \res_{z=c_\nu} \tr \Big(( Y^\nu)^{-1} \frac {d Y^\nu}{d z} \frac {\pa T^\nu}{\pa t^\nu_{ja} }  \Big)d z
&\& =- \res_{\zeta_\nu=0} \tr \Big(( Y^\nu)^{-1} \frac {d Y^\nu}{d \zeta_\nu} \frac {\pa T^\nu}{\pa t^\nu_{ja} }  \Big)d \zeta_\nu
\cr
=-\res_{\zeta_\nu=0} \tr \le(
 \frac {d T^\nu } {d \zeta_\nu} 
 \frac {\pa  T^\nu }{\pa t^\nu_{ja} }\ri)d \zeta_\nu
&\& -
\res_{\zeta_\nu=0} \tr \le(
\frac {d z}{d \zeta_\nu}(Y^\nu)^{-1} L Y^\nu
 \frac {\pa T^\nu }{\pa t^\nu_{j a} } \ri)d \zeta_\nu.
\eea
The first residue vanishes because  $\frac {\pa  T^\nu}{\pa t^\nu_{j a} } $ is a Laurent polynomial
containing only negative powers of $\zeta_\nu$.
The second residue written explicitly is
\be
-\res_{\zeta_\nu=0} \tr \le(
(Y^\nu)^{-1} L Y^\nu
\frac {E_{aa}} {j\zeta_\nu^j}\ri)\frac {d z}{d \zeta_\nu}d \zeta_\nu.
\ee
The proof will be complete if we can show that  the diagonal part of $(Y^\nu)^{-1}L Y^\nu\frac {d z}{d \zeta_\nu}$ coincides with the Taylor expansion 
of the eigenvalue matrix $\Lambda^\nu$ up terms of order $\mathcal O(\zeta_\nu^{d_\nu+1})$. 
For simplicity, consider a  pole at a finite $z=c_\nu$, so that $d z/d\zeta_\nu=1$ and let $d = d_\nu$. 
We then have 
 \be
 (Y^\nu)^{-1} L Y^\nu =\overbrace{ \frac {d T^\nu}{d\zeta_\nu}}^{\Lambda^\nu_-(\zeta_\nu)} + (Y^\nu)^{-1}  \frac {d Y^\nu}{d\zeta_\nu}.
 \ee 
 Let $\eta_+(\zeta_\nu)$ denote the diagonal part of $(Y^\nu)^{-1}  \frac {d Y^\nu}{d\zeta_\nu}$ and $F(\zeta_\nu)$ the off--diagonal part (both as formal analytic series). The goal is to show that $\eta_+$ coincides with the Taylor expansion of the eigenvalue matrix $\Lambda^\nu$ up to order $\zeta_\nu^{d+1}$.
 
 To see this, note that the spectrum of $(Y^\nu)^{-1} L Y^\nu$ coincides with that of $L$ and hence with the spectrum of 
 \be
 \Lambda^\nu_-(\zeta_\nu) + \eta_+(\zeta_\nu) + F(\zeta_\nu).
 \ee
 Define
 \be
 D := {\rm diag}(D_1,\dots, D_r)
 ,\ \ \ 
  D_a(\lambda, \zeta_\nu) := \lambda - (\Lambda^\nu_-(\zeta_\nu) + \eta_+(\zeta_\nu))_{aa}
  \ee
   and consider the characteristic polynomial 
 \be
 \det \Big(\lambda\Ib - \Lambda^\nu_-(\zeta_\nu) - \eta_+(\zeta_\nu) - F(\zeta_\nu)\Big) = \le(\prod_{a=1}^r D_{a}\ri) \det \le[\Ib - D^{-1} F\ri].
 \ee 
 Recalling that the diagonal elements of $F$ vanish, the expansion of the determinant has the form 
 \be
 \det \le[\Ib - D^{-1} F\ri]=1 + \mathcal O(\zeta_\nu^{2d_\nu+2})
 \ee
 near to $z-c_\nu$.
  The characteristic equation (\ref{char_eq_L}) implies that one of the $D_a(\lambda_a, \zeta_\nu)$'s vanishes to order $\zeta_\nu^{d+1}$, 
  \be
  \lambda_a = (\Lambda^\nu_- + \eta_+)_{aa} + \mathcal O(\zeta_\nu^{d_\nu+1}).
  \ee
  This concludes the proof of equality between \eqref{tnuja} and \eqref{H_t_nu_ja}.
  
For \eqref{cnu},  start by observing that 
\be
\pa_{c_\nu}T^\nu=-\pa_z T^\nu 
\ee
These equations hold for $\nu=1,\dots, N$, so $d z/d \zeta_\nu=1$,  and the RHS of  \eqref{cnu} is
\bea
&\& \res_{z=c_\nu} \tr \Big(( Y^\nu)^{-1} \frac {d Y^\nu}{d z} \frac {d T^\nu}{d z} \Big)d z
\cr
&\&= \res_{z=c_\nu} \tr \le(\frac{d T^\nu}{d z} \ri)^2  d z +
\res_{z=c_\nu} \tr \le((Y^\nu)^{-1} L Y^\nu \frac {d T^\nu }{d z } \ri)d z.
 \label{dcnu}
\eea
The first residue is again zero because the integrand is a negative  power Laurent polynomial starting with $(z-c_\nu)^{-2}$. We have already shown 
 that the diagonal part of $(Y^\nu)^{-1} L Y^\nu$ coincides with the diagonal matrix of eigenvalues $\Lambda^\nu$,  up to order $\mathcal O((z-c_\nu)^{d_\nu+1})$. 
Therefore,  in the second residue in \eqref{dcnu}, we can substitute $\Lambda^\nu$ for $(Y^\nu)^{-1}LY^\nu$. 
Since $d T^\nu/d z$ is the singular part of $\Lambda^\nu$ at $z=c_\nu$,  the second residue equals
\be
\res_{z=c_\nu} \Tr\big(\Lambda^\nu(z) \Lambda_{sing}(z)\big) d z=\frac 1 2  \res_{z=c_\nu} \Tr\big(\Lambda^\nu(z) \Lambda^\nu(z)\big) d z = \frac 1 2  \res_{z=c_\nu} \Tr\big(L^2(z)\big) d z.
\ee
We thus  have shown the equivalence of \eqref{tnuja} with  \eqref{H_t_nu_ja} and  \eqref{cnu} with  \eqref{H_c_nu}.
\end{proof}

%%%%%%%%%%%%%%%%%%% Section 4 Rational $L(z) %%%%%%
\section{Birkhoff connection, commutativity and Poisson property}
\label{Birkhoff_conn_commut_Poisson}

%%%%%%%%%%%%%%%%%%% Section 4.1 The case of polynomials $L(z)$} %%%%%%
\subsection{The Birkhoff connection}
\label{birkhoff_conn}
Consider the isomonodromic deformation equations  (\ref{t_nu_j_a_def_eq}), (\ref{c_nu_def_eq}) whose
compatibility conditions express the vanishing curvature  (\ref{zero_curv_eq}) of the connection $\Omega$ in \eqref{Omegadef}. 
In particular the Lax matrix $L(z)$ satisfies the following  zero curvature equations \cite{JMU}
\be
\frac{\pa L(z) }{\pa {t^\nu_{ja} }}  = \frac {d U^\nu_{ja }(z;L)}{d z}   - \Big[L(z), U^\nu_{ja}(z;L) \Big] ,\qquad
\frac{\pa L (z)}{\pa c_\nu}  = \frac {d V^\nu(z;L) }{d z}  - \Big[L(z), V^\nu(z;L) \Big] ,
\label{d41}
\ee
where the matrices $U^\nu_{ja}, V^\nu$ are defined in \eqref{defU},  \eqref{defV}. These equations should be viewed as defining commuting vector fields on the manifold $\LL_{r, \db}$ of rational matrices of the form \eqref{Lhat}, which is why we have  explicitly indicated the dependence of  
the connection component matrices $\{U^\nu_{ja}(z;L), V^\nu(z;L)\}$ on the Lax matrix $L$.

They can equivalently be written in terms of  {\it spectral} data. The following  is equivalent to eqs.~(\ref{defU}), (\ref{defV}):
\bea
U^\nu_{ja}(z; L)&\&:= \le( P^\nu(\zeta_\nu) \frac {\pa  T^\nu(\zeta_\nu)}{\pa t^\nu_{ja}} (P^\nu(\zeta_\nu))^{-1}\ri)_{sing}
= \le( P^\nu(\zeta_\nu) \frac { E_{aa} }{j \zeta_\nu^j } (P^\nu(\zeta_\nu))^{-1}\ri)_{sing}, 
\label{defU_spec}
\\
 \nu&\&=1,\dots, N, \infty
\cr
V^\nu(z; L)&\& :=  \le( P^\nu(\zeta_\nu) \frac {\pa  T^\nu(z)}{\pa c_\nu } (P^\nu(\zeta_\nu))^{-1}\ri)_{sing}
=- \le( P^\nu(\zeta_\nu) \frac {d T^{\nu}(z)}{d \zeta_\nu} (P^\nu(\zeta_\nu))^{-1}\ri)_{sing},
\label{defV_spec}
\\
 \nu&\&=1\dots, N
 \nonumber
\eea
The only difference is that we have replaced the formal series $Y^\nu(\zeta_\nu)$ in  \eqref{formal_Psi} with a local analytic series of eigenvectors. 
Note that right multiplication of the matrix of eigenvectors $P(z)$ by an invertible diagonal matrix
changes the normalization of the eigenvector, but does not affect  formulae \eqref{defU}, \eqref{defV}. 
The equivalence is due to the fact that 
\be
Y^\nu(\zeta_\nu) = P^\nu(\zeta_\nu) +\mathcal O(\zeta_\nu^{d_\nu+1}). 
\ee

Now we come to the crux of the matter. The vector fields defined by \eqref{d41} contain the sum of  two terms: a commutator term  
and the $z$-derivative of the deformation matrices appearing in eqs.~(\ref{t_nu_j_a_def_eq}). (\ref{c_nu_def_eq}).
The former has a clear interpretation. As will be proved explicitly in Section \ref{hamilt_vector fields_comm},
 the equations 
\be
\label{hamilvect}
\Xb_{H_{t^\nu_{ja}}} L:=\Big[U^\nu_{ja}, L\Big],\quad
\Xb_{H_{c_\nu}} L:=\Big[V^\nu, L\Big], \quad \Xb_{H^\infty_a} L:=\Big[E_{aa}, L\Big],
\ee
give the infinitesimal isospectral deformations  generated by the spectral invariant Hamiltonians  $\{H_{t^\nu_{ja}},  H_{c_\nu}, H^\infty_a\}$,
defined by the action of their Hamiltonian vector fields $\mathbf X_{H_{t^\nu_{ja}}}, \mathbf X_{H_{c_\nu}}, \Xb_{H^\infty_a}$.  Our focus however is on the other terms; namely,  the vector fields $\nabla_{t^\nu_{ja}}$ and $\nabla_{c^\nu}$ that act as follows on the matrix $L$ (i.e. on the linear functions of $L$)
\be
\label{birk1}
\nabla_{t^{\nu}_{ja}}L(z):=  \frac {d}{d z} U^\nu_{j  a }(z;L) ,\qquad
\nabla_{c^{\nu}}L(z):=  \frac {d}{d z} V^\nu (z;L)  
\ee
This is understood as defining the action of $\nabla_{t^\nu_{ja}}$ and $\nabla_{c^\nu}$ on each of the coefficients of $L$ (viewed as linear coordinates on $\LL_{r, \db}$) as well as the position of the poles, extended to arbitrary differentiable functions of $L$ by requiring it to be a derivation.
We would like to interpret  equations \eqref{birk1} as ``explicit derivatives" with respect to the parameters $\{t^\nu_{ja}, c_\nu\}$. 
But for this to make sense, we need to verify that:
\begin{enumerate}
\item 
The vector fields $\nabla_{t^{\nu}_{ja}}, \nabla_{c^\nu}$ commute for all $\nu, j$ and $ a$.
\item They act on the Casimir functions $\{t^\nu_{ja}, c_\nu\}$ as directional derivatives along coordinate curves.
\end{enumerate}
\begin{definition}
We  call the map that associates  the vector field $\nabla_t$ to $t\in \Tb$  as in \eqref{birk1} the {\bf Birkhoff connection}.
\end{definition}
The justification of the term ``connection'' will be given in Section \ref{Further},  where we interpret $\nabla$ as a flat Ehresmann  connection. 
In addition we will verify that these vector fields preserve the Poisson structure. 

We then have 
 \begin{theorem}
\label{thmcasi}
The vector fields $\nabla_{t^{\nu}_{ja}}, \nabla_{c^\nu}$ given by \eqref{birk1} act as follows on the Casimir functions $t^{\mu}_{kb}$, $c_\nu$:
\begin{subequations}
\be
 \nabla_{t^\nu_{ja}} c_\mu &=0
  \label{t_nu_c} 
  \\
\nabla_{c_\nu} c_\mu & = \delta_{\mu\nu}
 \label{c_nu_c_mu}
\\
\nabla_{t^{\nu}_{ja}}t^{\mu}_{kb} &= \delta_{\nu\mu}\delta_{ab} \delta_{jk}
\label{t_nu_t_mu}
\\
\nabla_{c^\nu} t^\mu_{ja}& =0,\qquad .
\label{c_nu_t}
\ee
\end{subequations}
That is, for any $t,s \in  \Tb$ we have 
\be
\nabla_s t  = \delta_{t,s}.
\ee
\end{theorem} 
\begin{proof}
To prove \eqref{t_nu_c},   \eqref{c_nu_c_mu} we first need to express $c_\nu$ as a function on ${\LL_{r, \db}}$. 
Take any of the entries of $L(z)$ that actually has a pole of order $d_\nu+1$ at $c_\nu$. We can then write 
\be
c_\nu =-\frac1{1+d_\nu}  \res_{z=c_\nu} z \frac {d}{d z} \ln L_{k\ell}(z) d z.
\label{c_nu}
\ee
The residue may be understood as a  contour integral on a  small circle around $z=c_\nu$, which remains constant under the deformations $ \nabla_{t^\nu_{ja}}, \nabla_{c_\mu}$. 

To prove \eqref{t_nu_c},  apply $\nabla_{t^\nu_{ja}}$ to \eqref{c_nu} and use the definition \eqref{birk1} of its action on $L_{k\ell}$, 
together with \eqref{defU_spec} to  obtain 
 \be
\nabla_{t^\nu_{ja}}c_\mu=-\frac 1{1+d_\mu}  \res_{z=c_\mu} z\frac {d}{d z} \le(\frac{\frac {d}{d z}(U^\nu_{ja})_{k\ell} (z)}{L_{k\ell}(z)}\ri) d z  = \frac 1{1+d_\mu}  \res_{z=c_\mu}  \le(\frac{\frac {d}{d z}(U^\nu_{ja})_{k\ell} (z)}{L_{k\ell}(z)}\ri) d z
\label{nabla_tnuja_c_nu}
\ee
where in the second equality we have used integration by parts. The numerator is either analytic at $z=c_\mu$ if $\nu\neq \mu$, or 
has a pole of order  at most $d_\nu+1$. In either case the ratio is analytic at $c_\mu$ and the residue is zero. 

To prove \eqref{c_nu_c_mu} we similarly use the fact that
\be
\nabla_{c_\mu} L_{ij}(z) = \frac {d}{d z}V^\mu_{ij}(z)
\ee
 is analytic at $z=c_\nu$.  We then have 
 \be
 \nabla_{c_\mu} c_\nu=0
 \ee
 for $\nu\neq \mu$.
   If  $\mu=\nu$, note that $V^\nu_{ij}(z)$ has a pole of order $d_\nu+2$ at $c_\nu$ and hence 
\be
\nabla_{c_\nu} c_\nu =-\frac
1{1+d_\nu}  \res_{z=c_\nu} z \frac {d}{d z}\le(\frac{\frac {d}{d z}V^\nu_{ij}(z)}{L_{ij}(z)}\ri) d z = \frac
1{1+d_\nu}  \res_{z=c_\nu} \le(\frac{\frac {d}{d z}V^\nu_{ij}(z)}{L_{ij}(z)}\ri) d z .
\ee
Recalling that $V^\nu$ is just the negative of the singular part of $L$ at $c_\nu$, in the above residue we can add 
the regular part without changing the value of the integral, to get 
\be
\nabla_{c_\nu}c_\nu=-\frac 1{1+d_\nu}  \res_{z=c_\nu} \le(\frac{\frac {d}{d z}L_{ij}(z)}{L_{ij}(z)}\ri) d z  = 1.
\label{nabla_c_nu_c_nu}
\ee

We  next prove \eqref{t_nu_t_mu}. For all $\mu=1,\dots, N,\infty$, the diagonalization  
\eqref{LP_Lambda_nu}, \eqref{LP_Lambda_inf}  can be written uniformly as 
\be
\frac {d z}{d \zeta_\mu} L(z) = P^{\mu}(\zeta_\mu)\Lambda^{\nu}(\zeta_\mu)  P^{\mu}(\zeta_\mu)^{-1},
\ee
where $\Lambda^\mu$ is the diagonal matrix whose entries consist of the Laurent series in $\lambda_a(\zeta_\mu)$ with a pole of order $d_\nu+1$.
Formulae\ \eqref{t_nu_ja_res}, \eqref{t_inf_ja_res}  are equivalent to the  matrix identities:
\be
T^{\mu}_k = -\res_{z=c_\mu}  {\zeta_\mu^{k}} P^{\mu}(\zeta_\mu)^{-1} L(z) P^{\mu}(\zeta_\mu)d z.
\ee
Applying the derivation $\nabla_{t^{\nu}_{j a}}$ to all the matrix components on both sides, using (\ref{t_nu_c}), we obtain:
\be
\nabla_{t^{\nu}_{ja}}T^{\mu}_k = -\res_{z=c_\mu} \le( P^{\mu}(\zeta_\mu)^{-1}\frac{d U^{\nu}_{ja}(z)}{d z}  P^{\mu}(\zeta_\mu)
-
\Big[(P^{\mu})^{-1}\nabla_{t^{\nu}_{ja}} P^{\mu}, \Lambda^\mu\Big]
\ri)\zeta_\mu^kd z.
\label{A19}
\ee
The commutator in \eqref{A19} is diagonal free and hence  we can discard it because the left side must be a diagonal matrix.
If $\mu\neq \nu$, the residue vanishes because the matrix $U^\nu_{ja}$ is analytic at $z=c_\mu$. If $\mu=\nu$ we  substitute 
the definition \eqref{defU} of $U^{\nu}_{j a}$  and simplify to obtain
\be
\nabla_{t^{\nu}_{ja}}T^{\nu}_k = -\res_{z=c_\nu} \le(( P^{\nu})^{-1}\frac{d}{d z} \le(P^{\nu} 
\frac{ E_{aa}}{j\zeta_\nu^j} (P^{\nu})^{-1}
\ri)_{\rm{sing}} P^{\nu}\ri)_D\zeta_\nu^kd z
\cr
= -\res_{\zeta_\nu=0} \le( (P^{\nu})^{-1}\frac{d}{d \zeta_\nu} \le(P^{\nu} \frac{ E_{aa}}{j\zeta_\nu^{j}} (P^{\nu})^{-1}
\ri)P^{\nu}\ri)_D\zeta_\nu^kd \zeta_\nu,
\ee
where  $(\cdot)_D$ denotes the diagonal part of the matrix $(\cdot)$, 
and we have added the regular part since it does not contribute to the residue. Finally, using the Leibniz rule we obtain
\be
\nabla_{t^{\nu}_{ja}}T^{\nu}_k 
=- \res_{\zeta_\nu=0} 
\le( 
\Big[(P^{\nu})^{-1} 
\frac{d  P^{\nu}}{d  \zeta_\nu},\frac{  E_{aa}}{j\zeta_\nu^j} \Big]   -  \frac{E_{aa}}{\zeta_\nu^{j+1}}
\ri)_D \zeta_\nu^kd \zeta_\nu = E_{aa} \delta_{j,k}
\ee
where, again, we have used the fact that  the commutator is diagonal free.
This proves eq.~(\ref{t_nu_t_mu}).

To prove  \eqref{c_nu_t}, we repeat the argument used in the first part of the proof,   with  $U^\nu_{ja}$. replaced by $V^\nu$.
(for $\nu= 1, \dots, N$.) We have 
\be
\frac {d T^\mu (\zeta_\mu) }{d \zeta_\mu}=- \res_{\xi_\mu=0}   P^{\mu}(\xi_\mu)^{-1} L(w) P^{\mu}(\xi_\mu)\frac{d \xi_\mu}{\zeta_\mu-\xi_\mu},
\label{dT_d_zeta}
\ee
where 
\be
\xi_\mu := w-c_\mu,\  \mu=1, \dots, N, \  \ \text{or } \ \xi_\infty = 1/w.
\ee
Acting with $\nabla_{c^\nu}$ and using (\ref{nabla_c_nu_c_nu}) for $\mu=\nu$, we get
\bea
\nabla_{c^\nu} \le( \frac {d T^\mu (\zeta_\mu)}{d \zeta_\mu}\ri) =
- \delta_{\mu\nu} \frac {d^2 T^\nu}{d \zeta_\mu^2} (\zeta_\mu) 
- \res_{\xi_\mu=0}   
\le(
(P^{\mu})^{-1} V^\nu P^{\mu}  -\Big[(P^\mu)^{-1} \nabla_{c^\nu} P^\mu,L\Big] \ri)  \frac{d \xi_\mu}{\zeta_\mu-\xi_\mu}
\label{nabla_nu_dT_mu_d_zeta_mu}
\eea
The commutator is again diagonal free and the first term in the residue  is analytic if $\nu\neq \mu$, so this vanishes.
For $\mu=\nu=1,\dots, N$ we get
\bea
\nabla_{c^\nu}&\&  \frac {d T^\nu}{d \zeta_\nu} =
-  \frac {d^2 T^\nu}{d \zeta_\nu^2}  
- \res_{\xi_\nu=0}   
\le(
(P^{\nu})^{-1}\frac {d V^\nu}{d \zeta} P^{\nu} \ri)_D  \frac{d \xi_\nu}{\xi_\nu-\zeta_\nu}
\cr
\mathop{=}^{\eqref{defU}, \eqref{defV}}&\&
-  \frac {d^2 T^\nu}{d \zeta_\nu^2}  + \res_{\xi_\nu=0}   
\le(
(P^{\nu})^{-1}
\frac {d}{d \zeta_\nu}\le( P^\nu(\xi_\nu) \frac {d T^{\nu}(\xi_\nu)}{d \xi_\nu} P^\nu(\xi_\nu)^{-1}\ri)
P^{\nu} \ri)_D  \frac{d \xi_\nu}{\zeta_\nu-\xi_\nu}
\label{nabla_nu_dT_nu_d_zeta_nu}
\eea
Integrating by parts and discarding the commutator term (which is diagonal free) we obtain
\be
\nabla_{c^\nu}  \frac {d T^\nu}{d \zeta_\nu} 
= -  \frac {d^2 T^\nu}{d \zeta_\nu^2} -
 \res_{\xi_\nu=0}   
\le(
 \frac {d T^{\nu}(\xi_\nu)}{d \xi_\nu} \ri)  \frac{d \xi_\nu}{(\xi_\nu-\zeta_\nu)^2} = -\frac {d^2 T^\nu}{d \zeta_\nu^2} + \frac {d^2 T^\nu}{d \zeta_\nu^2} = 0.
 \label{nabla_nu_dT_nu_d_zeta_nu2}
\ee
showing that $\nabla_{c^\nu}$ annihilates all the Birkhoff invariants.
\end{proof}

%%%%%%%%%%%%%%%%%%% Section 4.2 Commutativity of $\nabla$'s %%%%%%
\subsection{Commutativity of the vector fields $\nabla$}
\label{commutative_explicit_derivs}
 \begin{theorem}
\label{thmflat}
For all  $\mu, \nu=1,\dots, N$, and $\nu=\infty$,  the vector fields $\{\nabla_{c^\mu}, \nabla_{t^\nu_{ja}}\}_{ j=1,\dots d_\nu, a=1,\dots,r}$ commute amongst themselves.
\end{theorem} 
\label{section_commute}
We give the proof in two parts. First we show that the vector fields $\nabla_{t^\mu_{ja}}, \nabla_{t^\nu_{kb}}$ commute for $\mu\neq \nu$.
This follows from  the following Lemma \ref{lemmaortho}, which actually contains a stronger statement. 
We then localize the proof for a single $\nu$ and show that $\{\nabla_{c^\nu}, \nabla_{t^\nu_{ja}}\}$ 
commute for all pairs  $(j,a)$.
 
\begin{lemma}
\label{lemmaortho}
For $\mu\neq \nu$ the vector fields satisfy
\be
\label{A13}
\nabla_{t^{\nu}_{ja}} L^{\mu}_{k} = 0, \quad  \nabla_{c_\nu} L^{\mu}_k = 0,
\ee
and therefore
\be
\nabla_{t^{\nu}_{ja}} \nabla_{t^{\mu}_{kb}} L \equiv 0,  \quad  \nabla_{c^{\mu}}\nabla_{t^{\nu}_{ja}} L \equiv 0 , \quad \nabla_{c^{\nu}} 
\nabla_{c^{\mu}} \label{explicit_deriv_nu_j_a}L \equiv 0.
\ee
In particular, this implies that the vector fields $\nabla_{t^\nu_{ja}}$, $\nabla_{t^\mu_{kb}}$, $\nabla_{c_\nu}$  all commute for  $\mu\neq \nu$. 
\end{lemma}
 
\begin{proof}
Formula \eqref{A13} follows from the definition \eqref{birk1} of  $\nabla$.
For example the matrix $\nabla_{t^\nu_{ja}}L(z) = U^\nu_{ja}(z)$ (see \eqref{defU_spec}) is analytic at $z=c_\mu$ ($\nu\neq \mu$), which means that $\nabla_{t^\nu_{ja}}$ annihilates all entries of the coefficient  matrices $L^\mu_k$ in the polar expansion of $L$ near $z=c_\mu$, and similarly for $\nabla_{c_\nu}$. 

To prove  \eqref{explicit_deriv_nu_j_a}, note  that
\be
\nabla_{t^{\nu}_{ja}} \nabla_{t^{\mu}_{kb}} L(z) 
=\frac{d}{d z}\nabla_{t^{\nu}_{ja}} U^{\mu}_{kb} (z),
\ee
so it suffices to show that $\nabla_{t^{\nu}_{ja}} U^{\mu}_{kb} (z)$
 is  independent of $z$, with a similar statement for $V^\nu$. 
 To prove this, observe that $U^{\mu}_{kb}$ depends only on the Taylor expansion of $P^{\mu}(\zeta_\mu)$ up to order $d_\mu$. 
 As will be shown below, these coefficients, in turn,  depend only on the coefficients $L^{\mu}_{\{d_\mu+1, \dots, 1\}}$,  
 and therefore are annihilated by $\nabla_{t^{\nu}_{ja}}$.
   This implies that 
   \be
   \nabla_{t^{\nu}_{ja}} U^{\mu}_{kb} (z) \equiv 0.
   \ee
   
   To see this, we need to consider the equations defining the Taylor expansion of $P^\mu(\zeta_\mu)$.
   By comparing the series expansions of both sides of the eigenvector equation 
\be
   \frac {d z}{d \zeta} L(z) P^\mu (\zeta_\mu) = P^\mu(\zeta_\mu)\Lambda^\mu(\zeta_\mu),
\ee
it follows that we can always express the series $P^\mu(\zeta_\mu), \Lambda^\mu(\zeta_\mu)$ as 
\bea
P^\mu(\zeta_\mu)= P^\mu_0 + P^\mu_1 \zeta_\mu + P^\mu_2 \zeta_\mu^2 + \dots &\&= G^\mu \Big(\Ib + F^\mu_1 \zeta_\mu + F^\mu_2 \zeta_\mu^2 + \dots \Big), \\
\Lambda^\mu(\zeta_\mu)&\& = \sum_{j= -d_\mu-1} ^\infty\Lambda^\mu_{j}\zeta_\mu^j,
\eea
where the coefficients $F^\mu_j$ are {\it diagonal free} matrices and the $\Lambda^\mu_j$'s are purely diagonal. 
Setting $F^\nu_0:=\Ib$ and defining the coefficient matrices $\wh L^\nu_j$ by the identity
\be
(G^\nu)^{-1}L(z) G^\nu\frac {d z}{d \zeta_\mu} = \sum_{j=-d_\nu-1}^\infty \wh L_{-j}^\nu \zeta_\mu^j,
\ee
the matrices $P^\nu, \Lambda^\nu$ are  determined by the  recurrence relations
\bea
\label{ratlp}
\Lambda^{\nu}_{-d_\nu-1}&\&:= -T^{\nu}_{d_\nu} = \wh L^\nu_{d_\nu+1}\in \mathfrak h, \cr
F^{\nu}_\ell &\&:=- \ad_{T^{\nu}_{d_\nu}}^{-1} \le(\sum_{j= 1}^{\ell}
\Big( \wh L^{\nu}_{d_\nu+1-j} F^{\nu}_{\ell-j} -  F^{\nu}_{\ell-j} \L^{\nu}_{d_\nu+1-j}\Big)\ri),  \cr
\L^{\nu}_{-d_\nu-1+\ell} &\&:= \le(\sum_{j\geq 1}
\Big( \wh L^{\nu}_{d_\nu+1-j} F^{\nu}_{\ell-j} -  F^{\nu}_{\ell-j} \L^{\nu}_{-d_\nu-1+j}\Big)\ri)_D,\ \ \ \ell \geq 1.
\eea
From these recursive formulae  it follows that the matrix $P^\mu_\ell$ depends only on $L^{\nu}_{d_\nu+1}, \dots, L^{\nu}_{d_\nu+1-\ell}$.
This implies that  $U^\nu_{j a}$ (and similarly $V^\nu$) only depend only on the singular part of $L\frac {d z} {d \zeta_\nu}$ at the  point $z=c_\nu$, 
proving the claim that the coefficients in the Taylor expansion of $P^{\mu}$ up to order $d_\mu$ depend only on the 
coefficients $L^{\mu}_{\{d_\mu+1, \dots, 1\}}$.
\end{proof}

Formulae\ \eqref{ratlp}  show that the coefficients of $P^\nu(\zeta_\nu)$, and hence also of the matrices $U^\nu_{ja}, V^\nu$, depend polynomially on the coefficient matrices $L^\nu_j$ and are Laurent polynomials in the differences of the eigenvalues of the leading coefficient matrices $L^\nu_{d_\nu+1}$, 
and hence in the differences of the diagonal entries of $T^\nu_{d_\nu}$. We state this in the following.
 
\begin{corollary}
\label{corratl}
The matrices $U^\nu_{ja}(z;L), V^\nu(z;L), U^\infty_{ja}(z;L)$ depend polynomially on the entries of the coefficient matrices of $L$ and as Laurent polynomials in the differences of the eigenvalues $\{-t^\nu_{d_\nu,1}, \dots, -t^\nu_{d_\nu,r}\}$ of the leading order singularity 
matrices $L^\nu_{d_\nu+1}$.
\end{corollary}
 
\begin{proof}{(Of Theorem \ref{thmflat}.)}
We first prove the commutativity of the various vector fields $\nabla$ attached to a single pole. It is clear that the nature of the proof is entirely local,
so we will write  $P$ instead of $P^\nu(\zeta_\nu)$ for the local analytic series of eigenvectors, and $\Lambda$ for the  local
Laurent series $\Lambda^\nu(\zeta_\nu)$ in a punctured neighbourhood of $z=c_\nu$.
To reduce the number of indices, we denote the two vector fields 
\be
\nabla_1:= \nabla_{t^\nu_{k a}}, \ \ \nabla_2:= \nabla_{t^\nu_{\ell b}}
\label{nabla_1_2}
\ee 
and define correspondingly
\be
Q_1 := 
P \frac {\pa T^\nu} {\pa t^\nu_{ka}} P^{-1},\quad Q_2 := 
P \frac {\pa T^\nu} {\pa t^\nu_{\ell b}} P^{-1}, \ \quad 
U_1:= \Big(Q_1\Big)_{sing}\ \ \ U_2 := \Big(Q_2\Big)_{sing}
\label{d642}
\ee
which, more explicitly, are: 
\be
Q_1 = \frac{1}{k\zeta_\nu^k} P   E_{ a  a } P^{-1}, 
\quad
Q_2 = \frac1{\ell \zeta_\nu^\ell} P  E_{ b  b } P^{-1}.
\ee
 The theorem will follow if we  can show that 
\be
\nabla_{1} \nabla_{2} L(z) = 
\nabla_{2}\nabla_{1}  L(z),
\ee
where, by \eqref{birk1}, 
\be
\nabla_j L(z) := U_j(z)',
\ee 
with $U_1, U_2$  defined in \eqref{d642} and  $'$  denotes $\frac {d}{d z}$.
 Note that 
\bea
\nabla_2 U_1' = \bigg(\Big[\nabla_2 P P^{-1}, Q_1\Big] + P\nabla_2\nabla_1 \L P^{-1} \bigg)_{sing}' =  \bigg(\Big[\nabla_2 P P^{-1}, Q_1\Big] \bigg)_{sing}'.
\eea
where in the second equality we have used
$$
\nabla_2\nabla_1 \L
=\nabla_2 \le(-\frac 1{ \zeta^{k+1-2\epsilon}_\nu} E_{ a  a } + \mathcal O(\zeta^\epsilon)\ri) = \mathcal O(\zeta^\epsilon), 
$$ 
with $\epsilon=1$ if $\nu=\infty$ and $0$ otherwise. In either case, the term 
is locally analytic and hence drops out of the formula,
so it is sufficient to prove that
\be
 \frac{d}{d z} \Big[\nabla_2 P P^{-1}, Q_1\Big]_{sing} =\frac{d}{d z} \Big[\nabla_1 P P^{-1}, Q_2\Big]_{sing} .
\ee
The proof will be complete if we can show that:
\be
\label{dstronger}
\Big[\nabla_2 P P^{-1}, Q_1\Big]_{sing} = \Big[\nabla_1 P P^{-1}, Q_2\Big]_{sing} + \text{const}.
\ee
To do this, we introduce the following notation: given any matrix $M(z)$,  define $\wh M(z)$ to be 
\be
\label{hatnotation}
\wh M(z) := P \big(P^{-1} M P\big)_{OD} P^{-1} ,
\ee
where $(\cdot)_{OD}$ denotes the off-diagonal part the matrix $(\cdot)$. 
Thus, we consider $M$ modulo the commutant of $L$. 
Also define the inverse of the adjoint map
\be
\label{hatadL}
 \ad_L^{-1}  M(z)  := P \le( \ad_\Lambda^{-1}\big(P^{-1} M P\big)_{OD}\ri) P^{-1}  
\ee
where all terms are viewed as formal Laurent series near $z=c_\nu$.
Note that for any $\ad$--regular diagonal matrix $D = {\rm diag}( d_1,\dots, d_r)$ and any off-diagonal matrix $M$, the inverse of $\ad_D$ is well defined:
\be
(\ad_D^{-1} M\big)_{ij} = \frac {M_{ij}}{d_i-d_j}.
\ee
Definitions \eqref{hatnotation}, \eqref{hatadL} imply  
\be
[L(z),\ad_{L}^{-1} (M)] = \wh M(z).
\ee
By the definition (\ref{birk1}) of $\nabla$,
\be
\nabla_j  L(z) = \frac {d U_j}{d z} .
\label{d120}
\ee
On the other hand, by the Leibniz rule, we have
\be
\label{d119}
\nabla_j L(z)=
 \bigg[\nabla_j P P^{-1}, L\bigg]+ P \nabla_j \L P^{-1}. 
\ee 
Equating the two  and using  \eqref{hatnotation}, \eqref{hatadL}, we deduce  that 
\be
\label{d652}
\wh{\nabla_jP P^{-1}} =- \ad_L^{-1} (U_j').
\ee

We now prove the  identity \eqref{dstronger}. First,  we have 
\bea
\Big[\nabla_2 P P^{-1}, Q_1\Big]_{sing} =-\Big[\wh{\nabla_2 P P^{-1}}, Q_1\Big]_{sing}
\m{=}^{\eqref{d652}} \Big[\ad_L^{-1}(U_2') , Q_1\Big]_{sing}, 
\label{d134}
\eea
The first equality is due to the fact that $Q_1$ commutes with $L$.
Since $Q_{1,2}$ contains at most the power $\zeta_\nu^{-d_\nu}$ and $L \frac {d z}{d \zeta_\nu}$
 has a pole of order $d_\nu+1$ at $\zeta_\nu$, it follows that $\ad^{-1}_{L}$ decreases the power by
  $d_\nu+1$ (or $d_\infty-1$ if $\nu=\infty$). 
  Therefore we obtain (recalling $U_2 = (Q_2)_{sing}$) 
\be
\ad^{-1}_{L}\le(U_2'\ri) =  \ad^{-1}_{L}\le(Q_2'\ri) + \OO(\zeta^{d_\nu+\delta}_\nu),
\ee
where $\delta =-1$ for $\nu=1, \dots, N$ and $+1$ for $\nu=\infty$.

We can thus replace $U_2'$ by $Q_2'$ in \eqref{d134}  (within, at most, an additive constant,
 if $\nu=\infty$, $k=d_\infty$), to obtain
\bea
\label{d135}
\Big[\nabla_2 P P^{-1}, Q_1\Big]_{sing} =  \Big[\ad_L^{-1}\le( Q_2'\ri) , Q_1\Big]_{sing} +\text{const},
\eea
where
\bea
\label{d136}
\le[\ad_L^{-1}\le(Q_2' \ri) , Q_1\ri]_{sing} &\& =\le[\ad_L^{-1}\le(\le[P' P^{-1}, Q_2\ri] + P\pa_2 T' P^{-1}\ri) , Q_1\ri]_{sing}  \cr
 &\&=\le[ \ad_L^{-1}\le(\le[ P'P^{-1}, Q_2\ri] \ri) , Q_1\ri]_{sing}  .
\eea
In the last step, we have used the fact that the second term in the argument of $\ad_L^{-1}$ commutes with $L$  and hence drops out.

Consider now the quantity being projected on the principal  part and observe that it belongs to the 
range of $\ad_L$ since  $Q_1$ commutes with $L$.
(Recall that $\ad_L$ is an invertible map on matrices of this form.) Using 
\be
\ad_L Q_j=0,
\ee
 we get
\be
\ad_L\le[\ad_L^{-1}\Big([P'P^{-1}, Q_2] \Big) , Q_1\ri] 
\mathop{=}^{\text{\tiny (Jacobi + $[L, Q_1]=0$)}} \le[[P'P^{-1}, Q_2] , Q_1\ri] 
\mathop{=}^{\text{\tiny $([Q_1,Q_2]=0)$}}  \le[ Q_2 , [P'P^{-1},Q_1]\ri] ,
\ee
which  shows that 
\be
\label{d138} 
\le[\ad_L^{-1}\Big([P'P^{-1}, Q_2] \Big) , Q_1\ri]=\le[\ad_L^{-1}\Big([P'P^{-1}, Q_1] \Big) , Q_2\ri].
\ee
Combining these results, we can write the following chain of equalities, where above each 
 the relevant identity is indicated.
\bea
\Big[\nabla_2 P P^{-1}, Q_1\Big]_{sing} &\&\m{=}^{\eqref{d135}} \Big[\ad_L^{-1}(Q_2') , Q_1\Big]_{sing}  + \text{const}
\m{=}^{\eqref{d136}}\le[ \ad_L^{-1}\Big([P'P^{-1}, Q_2] \Big) , Q_1\ri]_{sing}  +\text{const} \cr
&\&\m{=}^{\eqref{d138} }\le[\ad_L^{-1}\Big([P'P^{-1}, Q_1] \Big) , Q_2\ri]
+\text{const}\m{=}^{\eqref{d136}\atop 1\leftrightarrow 2}  \Big[\ad_L^{-1}(Q_1') , Q_2\Big]_{sing} +\text{const} \cr
&\&\m{=}^{\eqref{d135}\atop 1\leftrightarrow 2}
\Big[\nabla_1 P P^{-1}, Q_2\Big]_{sing} +\text{const}.
\eea
This proves \eqref{dstronger} and hence completes the proof for this case. The case $\nabla_1=\nabla_{c^\nu}$ is proved similarly.
\end{proof}

%%%%%%%%%%%%%%%%%%% Section 4.3 Poisson preserving property %%%%%%
\subsection{Poisson preserving property of $\nabla$}
\label{section_Poisson}
Denote by $\PP$ the  bivector field  defining the Poisson bracket \eqref{split_rational_Rmatrix_PB}
\be
\{ f, g\}= \PP(df, dg).
\ee
Formulae \eqref{birk1} define the vector fields 
$\{\nabla_{t^{\nu}_{ja}}, \nabla_{c^\nu}\}$ on the finite-dimensional manifold ${\LL_{r, \db}}$ of rational matrices \eqref{Lhat},
 and we have proved that they commute and  act on the Casimir functions $\{t^{\nu}_{ja},c^\nu\}$ as we would expect from an ``explicit derivative". 
It remains to show that these fields are infinitesimal generators of Poisson morphisms; i.e., that 
\be
\LL_{\nabla_t}\PP=0,
\label{Lie_deriv_PP}
\ee
where $\Lie_{\nabla_t}$ denotes the Lie derivative with respect to $\nabla_t$ for  $t\in \Tb$. 
This follows from the fact that the $\nabla_{t^{\nu}_{ja}}, \nabla_{c^\nu}$'s are 
differences between Hamiltonian vector fields and  ``isomonodromic'' vector fields. The first automatically preserve the Poisson brackets, 
and the second do also (cf. e.g., Hitchin \cite{Hit} and Boalch \cite{Bo1}). However, it is quite straightforward to show this directly, 
which we now do.

 The Poisson invariance condition (\ref{Lie_deriv_PP}) is equivalent to the following.
  \begin{theorem}
\label{thmpoisson}
Let $t$ denote any of the isomonodromic times $t\in \Tb$ and  $\nabla_t$ be the corresponding  vector field. Then
\be
\label{NS}
\nabla_t\{f,g\} = \{\nabla_t f, g\}  + \{f, \nabla_t g\}.
\ee
In particular, if $f,g$ are in the joint kernel of all the $\nabla_t$'s,  their Poisson bracket
$\{f, g\}$ is also.
\end{theorem} 
\begin{proof}
It is sufficient to verify \eqref{NS} for any two linear functionals of $L$. We take linear functionals $\mathcal X, \mathcal Y$ of the form 
\bea
\XX(L):= \sum_\mu \res_{c_\mu} \tr \le(X_\mu (z)L(z)\ri)d z,\ \ \ X_\mu(z):= \sum_{j=\epsilon_\mu}^{d_\mu+\epsilon_\mu}  X_{\mu  j}\zeta_\mu^j\\
\YY(L):= \sum_\mu \res_{c_\mu} \tr \le(Y_\mu (z)L(z)\ri)d z,\ \ \ Y_\mu(z):= \sum_{j=\epsilon_\mu}^{d_\mu+\epsilon_\mu}  Y_{\mu  j}\zeta_\mu^j,
\eea
where $\epsilon_\mu=0$ for $\mu=1,\dots, N$ and $\epsilon_\infty=1$. 
Note that $\XX$ and $\YY$ do not depend on the positions of the poles, only on the matrix elements of the $L^\mu_j$'s. For example 
\be
\XX = \sum_{\mu} \sum_{j} \tr (X_{\mu  j} L^\mu_j).
\ee

Furthermore we have 
\be
\{\XX, \YY\} = \sum_\mu \res_{z=c_\mu} \tr \le(\Big[X_\mu(z), Y_\mu(z)\Big] L(z) \ri)d z 
\ee
For similar reasons, the bracket here is independent of  the pole loci. The identity is therefore
trivially satisfied by the $\nabla_{c_\nu}$'s and we focus only on the action of $\nabla_{t}$ where $t$ is one of the higher Birkhoff invariants $t^\nu_{ja}$. 
Writing $L= P^\mu \Lambda^\mu (P^\mu)^{-1}$ we have 
\bea
d \XX &\&= \sum_\mu\res_{z=c_\mu} \tr \bigg(X_\mu \big[d P^\mu( P^\mu)^{-1}, L\big] + X_\mu P^\mu d\Lambda^\mu (P^\mu)^{-1} \bigg)d z \cr
 &\& =  \sum_\mu\res_{z=c_\mu} \tr  \bigg( \big[ L, X_\mu \big] d P^\mu( P^\mu)^{-1}+ (P^\mu)^{-1} X_\mu P^\mu d\Lambda^\mu \bigg)d z.
 \label{dgenerald} 
\eea

We now need to compute $d (\nabla_t \mathcal X)$.  For $t$ equal to any of the parameters $\{t^\nu_{ja}\}$ denote,
for brevity,
\be
U_t := U^\nu_{ja}.
\ee
Then
\be
\nabla_t \left(\XX\right)= \sum_{\mu} \res_{z=c_\mu} \tr \bigg(\nabla_t (L X_\mu)\bigg)d z  = 
\sum_{\mu}\res_{z=c_\mu}\tr \bigg(U'_t X_\mu \bigg)d z  =-\sum_\mu \res_{z=c_\mu}\tr \bigg(U_t X_\mu'\bigg)d z.
\ee
Observe that, for $\mu\neq \nu$,  $U_t$ has a pole only at $c_\nu$ and is analytic at $c_\mu$  so the only residue that contributes is at $\nu=\mu$. In this case we have
\be
\nabla_t \left(\mathcal X\right)=- \res_{z=c_\nu}\tr \bigg(U_t X_\nu'\bigg)d z= - \res_{z=c_\nu}\tr \bigg(P^\nu \nabla_t T^\nu (P^\nu)^{-1} X_\nu'\bigg)d z,
\ee 
where in the last step we have added the negative part of 
\be
Q_t:= P^\nu \nabla_ t(T^\nu) (P^\nu)^{-1}
\ee
which does not contribute to the residue.

Now  take the differential of this function. Consider first the case when $t =t^\nu_{ja}$ and  observe that  
\be
\nabla_t T^\nu= \frac{ E_{aa}}{j\zeta_\nu^j},
\ee
 so that  
 \be d\nabla_t T^\nu = \frac {E_{aa}}{\zeta_\nu^{j+1}} d c_\nu.
 \ee
  Then 
\bea
d \left(\nabla_t \mathcal X\right) (L) = -\res_{z=c_\nu} \tr \bigg(\bigg[d P^\nu(P^\nu)^{-1},Q_t\bigg] X_\nu' +P^\nu\frac {E_{aa}}{\zeta_\nu^{j+1}} d c_\nu  (P^\nu)^{-1}X_\mu' \bigg) d z  =\cr 
=\res_{z=c_\nu}  \tr \bigg(d P^\nu(P^\nu)^{-1}\bigg[Q_t ,X_\nu'\bigg] +
 P^\nu\frac {E_{aa}}{\zeta_\nu^{j+1}} d c_\nu  (P^\nu)^{-1}X_\mu' 
\bigg) d z  
\eea
By inspection and comparison with \eqref{dgenerald}, we conclude that $d \nabla_t \mathcal X$ is given by 
\be
d \nabla_t \mathcal X = \ad_L^{-1} \le(\big[Q_t, X_\nu'\big]\ri) + \bigg((P^\nu)^{-1} X_\nu' P_\nu\bigg)_{aa} \frac{(P^\nu)^{-1}E_{aa}P^\nu}{\zeta_\nu^{j+1}} \d c_\nu.
\label{dungulate}
\ee
Note that $\ad_L^{-1}$ is well defined because $[Q_t, X']$ annihilates the commutant part of $X'$, given that  $Q_t=P \pa_tT P^{-1}$ 
belongs to the commutant subalgebra of $L$. 
We can now complete the computation
\bea
\le\{ \nabla_t \mathcal X, \mathcal Y\ri\}(L) &\&+ \le\{ \mathcal X, \nabla_t \mathcal Y\ri\}(L)=
\res_{z=c_\nu} \tr \bigg(L \bigg[d\nabla_t \mathcal X, d\mathcal Y \bigg]\bigg) d z 
+\res_{z=c_\nu} \tr \bigg(L \bigg[d \mathcal X, d\nabla_t\mathcal Y \bigg]\bigg) d z .\cr
&\&
\eea
Consider the first term and observe that in the first equality below we can drop the second term in  \eqref{dungulate}  from the trace, since it commutes with $L$ and (using the cyclicity of the trace) yields a vanishing contribution. Therefore
\bea
\res_{z=c_\nu} \tr \bigg(L \bigg[d\nabla_t \mathcal X, d\mathcal Y \bigg]\bigg) d z &\&
\m{=}^{\eqref{dungulate}}
\res_{z=c_\nu} \tr \bigg(L \bigg[\ad_L^{-1}\le(\big[Q_t, X_\nu'\big]\ri) , Y_\nu \bigg]\bigg) d z 
\nn\cr
=\res_{z=c_\nu} \tr \bigg(\bigg[L,  \ad_L^{-1}\le(\big[Q_t, X_\nu'\big]\ri) \bigg] Y_\nu \bigg) d z 
&\&=
\res_{z=c_\nu} \tr \bigg(\big[Q_t, X_\nu'\big]  Y_\nu \bigg) d z  = 
\res_{z=c_\nu} \tr \bigg(Q_t \big[Y_\nu, X_\nu'\big]   \bigg) d z.  \\
&\&
\eea
Repeating the computation for the second term we have 
\bea
\le\{ \nabla_t \mathcal X, \mathcal Y\ri\}(L) + \le\{ \mathcal X, \nabla_t \mathcal Y\ri\}(L) &\&
=\res_{z=c_\nu} \tr \bigg(Q_t \big[Y_\nu, X_\nu'\big]   \bigg) d z   + \res_{z=c_\nu} \tr \bigg(Q_t \big[Y_\nu', X_\nu\big]   \bigg) d z  \cr
&\&=\res_{z=c_\nu} \tr \bigg(Q_t\frac {d}{d z} \big[Y_\nu, X_\nu\big]   \bigg) d z  = 
\res_{z=c_\nu} \tr \bigg(Q'_t \big[X_\nu, Y_\nu\big]   \bigg) d z  \cr
&\&=\res_{z=c_\nu} \tr \bigg(U'_t \big[X_\nu, Y_\nu\big]   \bigg) d z  = \res_{z=c_\nu} \tr \bigg(\nabla_t L \big[X_\nu, Y_\nu\big]   \bigg) d z  \cr
&\&= \nabla_t\{\mathcal X, \mathcal Y\}(L), 
\eea
which completes the proof.
\end{proof}

%%%%%%%%%%%%%%%%%%% Section 4.4 Isospectral flows and their Hamiltonian formulation%%%%%%
\subsection{Isospectral flows and their Hamiltonian formulation}
\label{hamilt_vector fields_comm}

In this section we reprove, by direct  computation, that the Hamiltonian vector  fields  generated by the Hamiltonian functions $H^\nu_{ja} , H_{c_\nu}, H^\infty_a$,
defined in \eqref{spectral_Isomon_hamiltonians_t_nu}, \eqref{spectral_Isomon_hamiltonians_t_inf}, \eqref{SchlesHam}, \eqref{form_mon_infty} coincide with the vector fields $\mathbf X_{H_{t^\nu_{ja}}}, \mathbf X_{H_{c_\nu}}$,  $\mathbf X_{H^\infty_a}$ defined in \eqref{hamilvect}. 
For example this means that 
\be
\Big\{ L,H_{t^\nu_{kj}}(L) \Big\}  = \Big[U^\nu_{ja}, L\Big], 
\ee
where the Poisson bracket is as expressed in \eqref{split_rational_Rmatrix_PB}.
We recall that a compact way of expressing the Poisson bracket, using tensor product notation, is
\be
\{\m{L}^1(z), \m{L}^2(w)\}=\bigg[\Pi, \frac {\ds \m{L}^1(z) -\m{L}^1(w) }{z-w}\bigg],
\ee
where 
\bea
\Pi: \C^r\otimes \C^r &\&\to\C^r\otimes \C^r  \cr
\Pi(\un v\otimes \un w)&\& = \un w\otimes \un v
\eea
 is the order-reversing operator.
 
\begin{theorem}
\label{zero_curv_Lax}
The Hamiltonian vector fields on ${\LL_{r, \db}}$  corresponding to the spectral invariant Hamiltonians 
$H_{t^\nu_{j  a} }$, $H_{t^\infty_{j  a} }$, $H_{c_\nu} $,  and $H^\infty_a$ defined in 
(\ref{spectral_Isomon_hamiltonians_t_nu}), (\ref{spectral_Isomon_hamiltonians_t_inf}), \ref{SchlesHam}) and (\ref{form_mon_infty})
take the form:
\begin{subequations}
\bea
\Xb_{H_{t^\nu_{ja}}}L(z) =\Big\{ L(z), H_{t^\nu_{j  a} }\Big\}
&\&= \Big[ U^\nu _{ja }(z;L), L(z)\Big],
\label{U_t_comm} \\
\Xb_{H_{c_\nu}}L(z) = \Big\{L(z), H_{c_\nu}\Big\}
&\&= \Big[ V^\nu (z;L), L(z)\Big],
\label{V_nu_comm} \\
\Xb_{H^\infty_a}L(z) = \Big\{L(z), H^\infty_a \Big\}
&\&= \Big[ E_{aa}, L(z)\Big],
\label{E_aa_comm} 
\eea
\end{subequations}
where the matrices $\{U^\nu_{ja}, V^\nu\}$ are defined either by eqs.~(\ref{defU}), (\ref{defV}) in terms of the formal asymptotic expansions 
or, equivalently, by eqs.~(\ref{defU_spec}), (\ref{defV_spec}).  In the notation \eqref{Hamvec_commut}, \eqref{Rmatrix}, 
viewed as elements of the loop algebra $L\grgl(r)$ they  are equal to
\begin{subequations}
\bea
 U^\nu_{ja} &\&= -(d H_{t^\nu_{ja}})_-,\quad V^\nu = \  -(d H_{c_\nu})_-, \quad \nu=1, \dots, N, 
 \label{UV_dH_nu}
 \\
 U^\infty_{ja} &\&= (d H_{t^\infty_{ja}})_+,\ \,   \quad E_{aa} = (dH^\infty_a)_+. 
  \label{UV_dH_inf}
\eea
The Birkhoff invariants  $\{t^\nu_{ja}\}_{\nu=1, \dots, N, \, j=0,1,\dots, d_\nu, \, a=1, \dots, r}$ 
and $\{t^\infty_{ja}\}_{j=1,\dots, d_\infty, a=1, \dots, r}$   (i.e., including the exponents of formal monodromy at the finite poles 
but not those at $\infty$), are all Casimir elements of the Poisson bracket.

\end{subequations}
\end{theorem} 
\begin{proof}
From the equality between expressions  (\ref{tnuja}) and (\ref{H_t_nu_ja})  of Theorem \ref{d_log_tau_H} for $H^\nu_{ja}$
and the definition of the matrix of eigenvectors  
\be
L(z) = P^\nu(\zeta_\nu)\Lambda^\nu(\zeta_\nu) P^{\nu}(\zeta_\nu)^{-1} \ \text{ near to } z= c_\nu,
\ee
we have 
\be
H_{t^\nu_{j  a} }= -\res_{z=c_\nu} \tr \Big(L  P^\nu(\zeta_\nu)  \frac{1}{j\zeta_\nu^j}   E_{ a  a } (P^\nu(\zeta_\nu))^{-1}\Big)d z.
\ee 
We first compute the differential of $H^\nu_{j  a }$ on the manifold ${\LL_{r, \db}}$ . 
Denote by 
\be
M(z,\l) = \widetilde{ \l \Ib-L(z)}
\ee
 the classical adjoint of $\l \Ib-L(z)$.  For $(\l ,z)$ on the spectral curve  in a neighbourhood of a point $(\lambda_a(c_\nu), c_\nu)$ over $z=c_\nu$, 
\be
d \l_a  =\tr\le( \frac {M(z,\l_a ) dL(z)}{\tr M(z,\l_ a )} \ri) =\tr\Big( P^\nu(\zeta_\nu)  E_{ a  a } P^\nu(\zeta_\nu)^{-1} dL(z) \Big),
\ee
where we have used the fact that for a matrix $L$ with simple spectrum, the matrix 
\be
\frac {\widetilde{\l_a \Ib-L}}{\tr(\widetilde{\l_a \Ib-L})}
\ee
 is the spectral projector onto the $1$-dimensional eigenspace with the eigenvalue $\lambda_a$. 
Defining
\be
Q^\nu_{j  a }(\zeta_\nu) := P^\nu(\zeta_\nu)  \frac{ E_{ a  a } }{j\zeta_\nu^j} P^\nu(\zeta_\nu)^{-1},
\ee 
we have 
\be
d H_{t^\nu _{j a} } = -  \res_ {z=c_\nu}\tr\Big( d L\,Q^\nu_{j  a } + L\Big[ d P^\nu (P^\nu)^{-1}, Q^\nu_{j  a }\Big]\Big)d z
=- \res_ {z=c_\nu}\tr\Big( d L\,Q^\nu_{j  a }\Big)d z
\ee
where in the last equality we have used the cyclicity of the trace and the fact that $[L,Q^\nu_{j  a }]=0$. Therefore
\be
d H_{t^\nu_{j a} } =- \res_ {z=c_\nu}\frac {1}{j{\zeta_\nu}^j}\tr\Big( P^\nu  E_{
 a  a } (P^\nu)^{-1} d L(z) \Big)d z =  -\res_ {z=c_\nu}\tr\Big(Q^\nu_{j  a } (\zeta_\nu)d L(z) \Big)d z. 
 \label{pickup}
\ee
Now, for brevity, set $H= H_{t^\nu_{ja}}$ and $Q(z)= Q^\nu_{ja}(\zeta_\nu)$. (We write $Q(z)$ for simplicity, but keep in mind that 
this is a Laurent series centered at $z=c_\nu$, and for $\nu=\infty$ we set $c_\infty=\infty$.) We then have 
\bea
\Big\{L(z), H \Big\} &\&=-\res_{w=c_\nu} {\rm tr}_2 \Big\{ \m{L}^1(z), \m{L}^2(w)\Big\} \m{Q}^2(w)d w
=
-\res_{w=c_\nu} {\rm tr}_2 \le[\Pi, \frac{\ds \m{L}^1(z)-\m{L}^1(w)}{z-w}\ri] \m{Q}^2(w)d w
=
\cr
&\&
=\res_{w=c_\nu}\frac{[ Q(w), L(z)-L(w)]}{z-w}d w
=\le[ \res_{w=c_\nu}\frac{Q(w)d w}{z-w},  L(z) \ri],
\label{L_PB_H_t}
\eea
where we have again used the fact that  $[Q(w),L(w)]=0$ and that, for any pair of matrices $A, B$ 
\be
\tr_2\Big([\Pi, \overset{1}{A}]\, \overset{2} {B}\Big) = [A,B].
\ee

The residue in (\ref{L_PB_H_t}) produces precisely the singular part of $Q(w)$ at $z=c_\nu$, which completes 
the proof of the first equation in (\ref{UV_dH_nu}).

For $H_{c_\nu}=\frac 1 2\res_{z=c_\nu} \tr\Big(L^2(z)\Big)d z$ we have similarly
\be
d H_{c_\nu}=\res_{z=c_\nu} \tr\Big(L(z)d L(z)\Big)d z,
\ee
so that (with $H=H_{c_\nu}$) 
\bea
\Big\{L(z), H \Big\} &\&=\res_{w=c_\nu} {\rm tr}_2 \Big\{ \m{L}^1(z), \m{L}^2(w)\Big\} \m{L}^2(w)
=
\res_{w=c_\nu} {\rm tr}_2 \le[\Pi, \frac{\ds \m{L}^1(z)-\m{L}^1(w)}{z-w}\ri] \m{L}^2(w)
=
\cr
&\&
=\res_{w=c_\nu}\frac{[  L(z)-L(w),L(w)]}{z-w}
=\le[  - \res_{w=c_\nu}\frac {L(w)}{z-w},L(z)\ri].
\label{54}
\eea
The residue produces minus the singular part of $L(w)$ at $w=c_\nu$ (in the variable $z$) which is precisely the matrix $V^\nu(z)$, see \eqref{defU}, \eqref{defV}.

To prove (\ref{UV_dH_nu}) and (\ref{UV_dH_inf}), we need to identify $d H_t$ with an element of the co-tangent space to the submanifold ${\LL_{r, \db}}$, 
viewed as a Poisson submanifold of $L^*\mathfrak {gl}(r)$. Under the   pairing \eqref{trace_res_pairing},  $T^*{\LL_{r, \db}} $ is identified with 
the quotient of $L\mathfrak {gl}(r)$ by the ideal $\grI_{\cb, \db}$. We do that only for one case, leaving the remainder to the reader. 
Consider a finite pole $c_\nu, \nu\in \{1,\dots, N\}$ and one of the Hamiltonians $H_{t^\nu_{ja}}$. 

We pick up the computation from \eqref{pickup} and observe that the residue remains unchanged if we truncate $Q^\nu_{j a}$ modulo $\mathcal O(z-c_\nu)^{d_\nu+2}$.
Provisionally denote the resulting rational function, with only one pole at $z=c_\nu$ and a polynomial part at $\infty$, as $\wh Q^\nu_{ja}$,  We would like to use  Cauchy's theorem to deform the contour of integration from a small circle $|z-c_\nu|=\epsilon$ to a large circle $|z|=R$. However, in doing so, we would pick up the residues at all
the other poles of $d L(z)$. To prevent this, multiply $\wh Q^\nu_{ja}$ by a scalar polynomial $h(z)$ such that 
\be
h(z) = \le\{
\begin{array}{cc}
1 + \mathcal O(z-c_\nu)^{2d_\nu+2}, \  & z\to c_\nu,\\
\mathcal O( z-c_\mu)^{d_\mu+2},\ \ & z\to c_\mu, \ \ \mu\neq \nu.
\end{array}
\ri.
\ee
We can then replace $ Q^\nu_{ja}$ in the residue \eqref{pickup} with $h(z) \wh Q^\nu_{ja}(z)$ without affecting the value of the residue. 
But now we can use Cauchy's theorem, since $d L(z) h(z) \wh Q^\nu_{ja}(z)$ has only one finite pole at $z=c_\nu$. 
Thus we have shown that $d H^\nu_{ja} $ can be identified with $ - h(z) \wh Q^\nu_{ja}(z)$ in the cotangent space, where the minus sign is due to the sign in front of the residue \eqref{pickup}.
Finally, from the properties of $h$, it follows that the projection of $ h(z) \wh Q^\nu_{ja}(z)$ in $L_-\mathfrak {gl}(r)$ 
is exactly the same as the principal part of $Q^\nu_{ja}$ at $z=c_\nu$. That is,
\be
R_0(d H^\nu_{ja}) = -\le(-h(z)\wh Q^\nu_{ja}\ri)_- = \big(Q^\nu_{ja}\big)_{sing} = U^\nu_{ja}.
\ee
For the case $\nu=\infty$, since the residue formula for $H_{t^\infty_{ja}}$ involves {\it minus} the residue at $\infty$  (which is a {\it positively} oriented contour integral along a circle), we find that $d H_{t^\infty_{ja}}$ is identified simply with the Laurent expansion of $Q^\infty_{ja}(z)$ ad then 
\be
R_1(d H_{t^\infty_{ja}}) = \le(Q^\infty_{ja}\ri)_+ = U^\infty_{ja}
\ee

To prove that the Birkhoff invariants are Casimir elements we proceed similarly. Consider $t^\nu_{ja}$ as in \eqref{princ_part_coeffs}, \eqref{form_mon_nu}. 
(This computation includes the exponents of formal monodromy at the finite poles.) Then 
\be
\d t^\nu_{ja} = -\res_{z=c_\nu} \tr\Big(W(z) dL(z) \Big)d z , \ \  W:= \ \zeta_\nu^j P^\nu(\zeta_\nu)  E_{ a  a } P^\nu(\zeta_\nu)^{-1}.
\ee
We now proceed similarly to the proof of \eqref{L_PB_H_t}:
\bea
\Big\{L(z), t^\nu_{ja} \Big\} &\&=-\res_{w=c_\nu} {\rm tr}_2 \Big\{ \m{L}^1(z), \m{L}^2(w)\Big\} \m{W}^2(w)d w
=
-\res_{w=c_\nu} {\rm tr}_2 \le[\Pi, \frac{\ds \m{L}^1(z)-\m{L}^1(w)}{z-w}\ri] \m{W}^2(w)d w
=
\cr
&\&
=\res_{w=c_\nu}\frac{[ W(w), L(z)-L(w)]}{z-w}d w
=\le[ \res_{w=c_\nu}\frac{W(w)d w}{z-w},  L(z) \ri]
\label{577}
\eea
Since $W(w)$ is locally analytic at $z=c_\nu$, the  residue in \eqref{577} is the null matrix and the Poisson bracket vanishes 
for all $t^{\nu}_{ja}$, which are therefore all Casimir elements. 

Note  that for $\nu=\infty$ and $j=0$ we have $t^\infty_{0a} = H^\infty_a$,  and these are not Casimir elements. However the same computation shows that in this case  the residue equals the constant matrix $E_{aa}$,  proving \eqref{E_aa_comm} and showing that the exponents of formal 
monodromy at $\infty$ are indeed the Hamiltonian generators of the group of invertible diagonal constant matrices, acting by conjugation on $L$.
\end{proof}

%%%%%%%%%%%% Section 5. Further discussion and open problems %%%%%%
\section{Further discussion and open problems}
\label{Further}

%%%%%%%%%%%% Subsection 5.1 Birkhoff fibration %%%%%%%%%%
\subsection{Birkhoff fibration}
\label{birkhoff_fibration}

To further clarify the meaning of the Birkhoff  connection, denote the space of loci of the finite, distinct  poles of the rational Lax matrices  in {$\LL_{r, \db}$ by
 \be
  \C^N_{\Delta}:= \{\cb:=(c_1,\dots, c_N),\  \{c_\nu \neq c_\mu,    \text{ for } \nu\neq \mu\} ,
 \ee
 the space of  $(\sum_{\nu=1}^N d_\nu +d_\infty)$-tuples of  diagonal $r \times r$ matrices by
 \be
 \prod_{\nu=1..N,\infty} (\grh)^{d_\nu-1}\times \grh_{reg}  := \{T^\nu_j\in \mathfrak h\}_{ j=1\dots d_\nu, \ \nu = 1, \dots, N, \infty}, 
 \text{ with } T^\nu_{d_\nu} \in \mathfrak h_{reg},
 \ee
  where $T^\nu_{d_\nu}\in \mathfrak h_{reg}$ has distinct eigenvalues, and the Cartesian product of these by
\be
\TT:=\C^N_{\Delta}\times\prod_{\nu=1..N,\infty} (\grh)^{d_\nu-1}\times \grh_{reg} .
\ee
The map $\Phi:{\LL_{r, \db}} \to \TT$ that associates to each $L\in \LL_{r,\db}$ the loci of its poles and the (higher) Birkhoff invariants is surjective, and the fibers are the union of the symplectic leaves over all values of the residual Casimir invariants $(T_0^1,\dots, T_0^N)\in  (\grh)^{N}$. 
If  all $d_\nu$'s are assumed $>1$, this realizes $\LL_{r, \db}$ within the open dense stratum of generic symplectic leaves as a fiber bundle over $\TT$, 
with fibers isomorphic to the space  given by
\be
\Phi^{-1}(\{\cb, {\bf t}\}) = 
\le\{P^\infty(\zeta) =\le(\Ib + \sum_{j=1}^{d_\infty-1} F^\infty_j\zeta^j\ri)\in \grGl(r)[[\zeta]]/{\rm Diag}(r)[[\zeta]]\mod \zeta^{d_\infty}\ri\} \cr
\times\prod_{j=1}^N \le\{P^\nu(\zeta) = G^\nu\le(\Ib + \sum_{j=1}^{d_\nu} F^\nu_j\zeta^j\ri)\in \grGl(r)[[\zeta]]/{\rm Diag}(r)[[\zeta]]\mod \zeta^{d_\nu+1}\ri\}\times  {\mathfrak h}^N, 
\ee
where ${\rm Diag}(r)[[\zeta]]$ denotes the subgroup of formal series of  $r \times r$ diagonal matrices 
\be
D(\zeta)= D_0 + D_1\zeta + \dots, \quad \det D_0\neq 0,
\ee
 acting by right multiplication.  (If some of the $d_\nu=0$, then as many terms in the last factor $\mathfrak h^N$ should be replaced by $\mathfrak h_{reg}$).

Given $( \cb, \tb )\in \TT$ and  an element  $\le(P^\infty, P^1,\dots, P^N)\times (T^1_0,\dots, T^N_0\ri)$, we recover the matrix $L(z)\in \LL_{r, \db}$ 
from the formula 
\be
L(z) = &\le(P^\infty\le(\frac 1 z\ri) \le(\sum_{j=1}^{d_\infty} T^\infty_j z^{j-1}\ri) P^\infty\le(\frac 1 z \ri)^{-1}\ri)_{+} +\cr
\\ &+ \sum_{\nu =1}^N\sum_{j=1}^N \le(
P^\nu(\zeta_\nu)\le(-\sum_{j=1}^{d_\nu} \frac{T^\nu_j}{(z-c_\nu)^{j+1}}  +\frac {T^\nu_0}{z-c_\nu}\ri)P^\nu(\zeta_\nu)^{-1}\ri)_{sing}.
\ee

Thus $\frac{\partial}{\partial t} \ra \nabla_t$ lifts the tangent vectors from $T \TT$ to $T {\LL_{r, \db}}$ as a flat connection preserving the Poisson structure.

%%%%%%%%%%%%%%%%%%% Section 5.2 Quotient manifold and deautonomization%%%%%%
\subsection{Quotient manifold and deautonomization}

Let 
\be
\mathfrak T:= \Span\{\nabla_t, \ t \in \Tb\}.
\ee
 
 \begin{proposition}
\label{constrank}
 $\mathfrak T$ is an integrable distribution of constant, maximal rank
\be
N+ r \sum_{\nu=1}^N  d_\nu  + r d_\infty, 
 \ee
   and the  canonical projection $\pi:{\LL_{r, \db}} \to \mathcal W = {\LL_{r, \db}}/\mathfrak T$ is Poisson.
\end{proposition}
 
\begin{proof}
The integrability follows  from Theorem \ref{thmflat}. The rank condition is almost obvious, but we provide a proof nevertheless.
Suppose, on the contrary,  that there exists $L_0(z)\in {\LL_{r, \db}}$ such that  that the vector fields  $\nabla_{t^\nu_{j a}}$ are linearly dependent in  $T_{L_0}{\LL_{r, \db}}$. This implies that there  are Laurent polynomial diagonal  matrices $ D_\nu(\zeta_\nu)$ with poles of degrees $\leq d_\nu$  such that 
\be
\sum_{\nu =1, \dots , N, \infty} \frac {d }{d z}\Big(P^\nu(\zeta_\nu) D_\nu(\zeta_\nu) (P^\nu)^{-1}(\zeta_\nu)\Big)_{sing} \equiv 0.
\label{d510}
\ee 
Since each term in the sum is rational, with a single pole at $z=c_\nu$,  they must separately vanish. By definition of the projection $(\, \cdot \,)_{sing}$ at the finite poles $c_\nu,\, \nu=1,\dots, N$, the terms are Laurent polynomials  without constant term, but then $D_\nu(\zeta_\nu)\equiv 0$ (since the projection cannot result in a constant).
For $\nu=\infty$,  eq.~\eqref{d510} implies  that $(P^\infty D_\infty(z) (P^\infty)^{-1})_{sing}$ should be a constant, which is possible only if $D_\infty$ is a constant diagonal matrix. However this is not possible  because $\nabla  T^\infty$ is either a polynomial without constant coefficient or zero. 

Now consider the second statement. This is equivalent to saying that the algebra of $\nabla_t$ invariant functions is a Poisson subalgebra, which follows 
from Theorem \ref{thmpoisson}. 
\end{proof}

The dimension of $\mathfrak T$  is 
\be
\dim \mathfrak T =N+  r d_\infty + r\sum_{\nu=1}^N d_\nu,
\ee
while the dimension of ${\LL_{r, \db}}$ is 
\be
\dim {\LL_{r, \db}} = N + r + r^2(d_\infty-1) + r^2 \sum_{\nu=1}^N (d_\nu+1). 
\ee
Recall that the fibers of the projection ${\LL_{r, \db}} \to \TT$ are {\it unions} of symplectic leaves over the $rN$ values of the Casimir functions
$\{t^\nu_{0a}\}_{\nu=1, \dots, N \atop  a=1, \dots r}$ (the exponents of formal monodromy). 
There are an additional 
\bea
2K + rN &\&:= \dim {\LL_{r, \db}}- \dim \mathfrak T = r +r^2 (d_\infty-1 )+ r^2 \sum_{\nu=1}^N  d_\nu
-r d_\infty  - r\sum_{\nu=1}^N  (d_\nu+1)  \cr
&\&=  r(r-1)\left(d_\infty +\sum_{\nu=1}^N d_\nu +N -1\right) +rN.
\eea
functionally independent $\nabla$ invariants. These include the $rN$ exponents of formal 
monodromy$\{t^\nu_{0a}\}_{\nu=1,\dots, N, \infty, \, a=1, \dots ,r}$  at the finite poles,
 which we denote  as
\be
\TT_0:=\{t^\nu_{0a}, \ t^\nu_{0a}\neq t^\nu_{0b} \ \text{ if } d_\nu\neq 0 \text{ and } a\neq b\}_{\nu=1, \dots N, \,  a, b=1, \dots r} .
\ee
Setting  both these and all elements of $\TT$ equal to constants, we obtain the symplectic leaves, which are of dimension $2K$.
 We may therefore choose a complementary set of coordinates consisting of $\nabla$-invariant coordinates on the symplectic
 leaves, which we denote
\be
\Wb := (w_1,\dots, w_{2K}).
\ee 
Using the Riemann-Hurwitz formula, we can verify that the dimension  $2K$of the  symplectic leaf is 
related to the genus $g$ of generic spectral curves $\CC$ by
\be
K =g+r-1.
\ee
We thus have, at least locally,  a full set of complementary (holomorphic) functions 
$\{w_\alpha, \ t^\nu_{0a}\}_{{\alpha =1, \dots, 2K \atop \nu=1, \dots, N} \atop  a=1, \dots r} $ 
which,  together with the isomonodromic deformation parameters $\{t^\nu_{j a}, c_\nu\}_{{\nu=1, \dots, N, \infty \atop \ a=1, \dots, r} \atop j=1, \dots d_\nu}$,
form a  (local) coordinate system on $\LL_{r, \db}$. Note that the first $r$ of these $(w_1, \dots w_r)$ may be chosen to be the 
exponents of formal monodromy $\{t^\infty_{0a}=H^\infty_a\}_{a=1, \dots, r}$ at $\infty$, which all Poisson commute amongst themselves,
and with all the spectral invariants $\{H^\nu_{ja}\}$, and these can be completed to form Darboux coordinate systems, at least locally, on the symplectic leaves.

%%%%%%%%%%%%%%%%  Subsection 5.3 Isomonodromic equations as deautonomization of isospectral deformations. %%%%
\subsection{Isomonodromic deformations as deautonomization of isospectral flows.}
\label{deautonom_isomon}

Suppose  that we have found a full set of $2K+rN$ independent $\nabla$-invariants $(\Wb,\TT_0)$, 
including  the  exponents of formal monodromy $\TT_0$. 

If $w$ is a $\nabla$--invariant function  then the isomonodromic equations read 
\be
\frac{\pa}{\pa t} w = \underbrace{\nabla_t w }_{=0}+ \{w, H_t\},\ \ \ \forall \, t\in \Tb, 
\ee
with the Hamiltonians $H_t$ defined in \eqref{spectral_Isomon_hamiltonians_t_nu}, \eqref{spectral_Isomon_hamiltonians_t_inf},  \eqref{SchlesHam}.
This means that  the isomonodromic equations would then take standard non-autonomous Hamiltonian form
(since the Hamiltonians also depend explicitly on the isomonodromic times). 

The question that remains is  how to determine such invariants explicitly  and  use them to parametrize the Lax matrix $L(z)$ canonically.
This means completing the transverse Birkhoff invariants and pole loci with $\nabla$  invariants so as to form a (local) coordinate system on
the phase space, which is canonical on the symplectic leaves, thereby simultaneously implementing the 
transversal foliation and the symplectic one as a local product.  

This has been resolved in special cases in the literature; e.g., in \cite{HR} for the six Painlev\'e\ equations.
In \cite{HTW} it was done for extensions of the $P_{II}$ system to systems of isomonodromic deformation
equations in which the  $\grs\grl(2)$-valued Lax matrix consists of the sum of  a  first degree polynomial part  plus any
number first order poles. In \cite{MaMo} it was done for the $P_{II}$  hierarchy, represented as zero curvature equations
involving $\grs\grl(2)$-valued  Lax matrices $L(z)$ that are polynomials  in $z$ plus a fixed first order pole at $z=0$, 
that satisfy the involutive symmetry 
\be
L(-z)=\sigma_1L(z)\sigma_1.
\ee 
The system was shown to consist of nonautonomous deformations of Hamiltonian equations with respect to
the rational classical $R$-matrix structure on loop algebras, with the Lax matrix parametrized by a combination
of the Casimir invariants of higher Birkhoff type, and $\nabla$--invariants that are chosen to form a Darboux
coordinate system.

It is not evident how to generalize these constructions even to  arbitrary polynomial $\grs\grl(2)$-valued  $L(z)$.
A further example, however, given in Appendix \ref{example_A}, does this  for the case of cubic polynomials, 
suggesting that it may be possible for all $\grs\grl(2)$-valued polynomial $L(z)$'s, and
 perhaps more generally, for all rational ones.
 
\appendix
\label{App_A}
\renewcommand{\theequation}{\Alph{section}.\arabic{equation}}

\section{Example:  Traceless polynomial $L(z)$ with $r=2, \ d_\infty=4$}
\label{example_A}
Consider the case of  a polynomial traceless Lax matrix $L(z)\in\mathfrak {sl}_2$ with $d_\infty=4$. Since the traces are all Casimir elements, 
 the above proofs  hold when $L(z)$ is traceless; only the dimension count is slightly different.
 An explicit parametrization in terms of canonical coordinates is given as follows:
\bea
L &\&=
\left( 
t_{4}\,{z}^{3}+t_{3}\,{z}^{2}+ \left( t_2-\sqrt {t_{4}}y_{2}\,x_{1}
 \right) z-\sqrt [4]{t_{4}}\le(x_{1}\,y_{3}+x_{3}\,y_{2}\ri)+t
_{1}-{\frac {t_{3}\,x_{1}\,y_{2}}{3\sqrt {t_{4}}}}
\right) \sigma_3\cr
&\&+  \sqrt{2}\left(
-{t_{4}}^{3/4}x_{1}\,{z}^{2}
-\left(
	 \sqrt {t_{4}}x_{3}
	 +{\frac {2t_{3}x_{1}}{3\sqrt [4]{t_{4}}}} 
	 \right) z
+\frac 1 4\sqrt [4]{t_{4}}{x_{1}}^{2}y_{2}
-{\frac {t_{3}x_{3}}{3\sqrt {t_{4}}}}
-\sqrt [4]{t_{4}}x_{2}
-{\frac {t_{2}\,x_{1}}{2\sqrt [4]{t_{4}}}}
+{\frac {{t_{3}}^{2}x_{1}}{18{t_{4}}^{5/4}}}
\right)\sigma_+ \cr
%%%%%%%%%%%
&\& + \sqrt{2} \left(
-{t_{4}}^{3/4}y_{2}{z}^{2}
- \left(
	 \sqrt {t_{4}}y_{3}
	 +{\frac {2t_{3}y_{2}}{3\sqrt [4]{t_{4}}}} 
\right) z
+\frac 1 4\sqrt [4]{t_{4}}{y_{2}}^{2}x_{1}
-{\frac {t_{3}\,y_{3}}{3\sqrt {t_{4}}}}
-\sqrt [4]{t_{4}}y_{1}
-{\frac {t_{2} y_{2}}{2\sqrt [4]{t_{4}}}}
+{\frac {{t_{3}}^{2} y_{2}}{18{t_{4}}^{5/4}}}
\right)\sigma_-, \cr
&\&
\label{cubic_Lax}
\eea
where
\be
t_j = t^{\infty}_{j 1} = - t^{\infty}_{j 2}, \quad j=1, \cdots ,4
\ee
are the Casimir elements defined in  \eqref{princ_part_coeffs_inf}, and $(x_1, x_2, x_3, y_1, y_2, y_3)$ are Darboux coordinates
on the symplectic leaves.
\be
\{x_i, x_j\}= \{y_i, y_j\} =0, \quad \{x_i, y_j \} = \delta_{ij}, \quad i,j, = 1, 2, 3.
\ee
(These coordinates were found by an explicit computation of the $\nabla_{t_i}$--invariants using a computer algebra system.)

The exponent of formal monodromy is 
\be
\label{t0theta}
t^\infty_0 := -\res_{z=\infty} \sqrt{-\det L(z)} d z= a := x_1y_1 + x_2 y_2 + x_3y_3, 
\ee
which is a constant of motion that Poisson commutes with all other spectral invariants, but is not a Casimir invariant.
The Hamiltonian vector field corresponding to $a$ is given by the commutator
\be
\Xb_a(L) = \{L, a\} = [\sigma_3, L]
\ee
 and the flow it generates is the group of  scaling transformations, given by conjugation with invertible diagonal matrices
\bea
f_s: \{x_i, y_i, t_a\} &\& \ra \{e^s x_i, e^{-s} y_i, t_a\} , \quad i=1,2, 3, \quad a = 1,2,3,4 ,
\label{scaling}
\\
f_s: L(z) &\&\ra e^{s\sigma_3}L(z) e^{-s\sigma_3}.
\label{scaling_Lax}
\eea

It follows by  direct computation that 
\be
\pa_{t_j} L(z) = U_j'(z) = \frac \d{d z}\le(\frac {z^j}jP \s_3 P^{-1}\ri)_+, \quad j=1,\dots, 4.
\ee
The affine symmetry group of dilations and translations in $z$ acts  on the Lax matrix $L(z)$ as follows
\bea
c \wt L(c z + b, \wt {\bf t} ) &\& =  L(z, {\bf t}),  \\
(\wt t_1, \wt t_2, \wt t_3, \wt t_4) &\& := (c(t_1 + bt_2 + b^2 t_3 + b^3 t_4), c^2 (t_2 + 2bt_3 + 3b^2t_4) , c^3( t_3 + 3 t_4  b), c^4 t_4).
\eea
Choosing $c = t_4^{-\frac 1 4}$ and $b = - \frac {t_3} {3 t_4}$, this   has the same  effect as setting $t_4=1, \ t_3=0$,  reducing $L(z)$ to
\bea
L(z) &=\left({z}^{3}+ \left(t_2-x_{1}\,y_{2} \right) z-x_{1}\,y_{3}-x_{3}\,y_{2}+t_{1} \right) \sigma_3\cr
&- \sqrt {2}\left( x_{1}\left( {z}^{2}+\frac{t_{2}} 2 \right) 
+x_{3}\,z+x_{2}  -\frac 1 4 y_{2}\,{x_{1}}^{2}\right) \sigma_+ \cr
& -\sqrt {2} \left( y_{2}\, \left( {z}^{2}+\frac{t_{2}}2 \right) +y_{3}\,z+y_{1}  -\frac 1 4x_{1}\,{y_{2}}^{2} \right) \sigma_-.
\label{cubic_Lax_simplif}
\eea
There are therefore only two relevant isomonodromic deformation parameters, $(t_1,t_2)$.
 Expanding the eigenvalue $\lambda = \sqrt{-\det L}$ (on one of the sheets) 
\be
\lambda = z^3 + t_2z  + t_1 + \frac {a}{z} + \frac {H_1}{2z^2} + \frac {H_2}{2z^3} + \mathcal O(z^{-4}) ,
\ee
where 
\bea
H_1 =&\& \frac 3 2{x_{1}}^{2}y_{2}y_{3}+ \left(\frac  3 2 x_{3}{y_{2}}^{2}-
2t_{1}y_{2}-
t_{2}y_{3} \right) x_{1}
+ 2 x_{2}y_{3}+ \left(2y_{1} -y_{2} t_{2} \right) x_{3}
\label{Ham_1}
\\
H_2 =&\& \frac 1{16} {x_{1}}^{3}{y_{2}}^{3}+ 
\left( 
\frac 1 2 {y_{3}}^{2}-\frac 1 4 t_{2}{y_{2}}^{2}-\frac 5 4 y_{1}y_{2}
 \right) {x_{1}}^{2}+ 
 \left( \frac 1 2t_{2}y_{1}
+ \left( \frac 1 4{t_{2}}^{2}+{t_{0}} \right) y_{2} -\frac 5 4 x_{2}{y_{2}}^{2}-t_{1}y_{3} \right) x_{1}
\cr
&\&+ \left( \frac 1 2 y_{2}t_{2}+y_{1}
 \right) x_{2}+\frac 1 2 {x_{3}}^{2}{y_{2}}^{2}-t_{1}x_{3}y_{2}-  t_{2}a.
 \label{Ham_2}
\eea
are the nonautonomous Hamiltonians $H_1:=H_{t^\infty_{11}}, \ H_2 := H_{t^\infty_{21}}$  obtained from  
formulae\ \eqref{spectral_Isomon_hamiltonians_t_inf},
and
\be
a = H^\infty_1 = -H^\infty_2
\label{a_H_inf}
\ee
is the exponent of formal monodromy at $z=\infty$.
The deformation matrices $U_1,U_2$ then become 
\bea
U_1 = \left[ \begin {array}{cc} z&-\sqrt {2}x_{1}\\ \noalign{\medskip}-
\sqrt {2}y_{2}&-z\end {array} \right],\qquad
 U_2 = \frac 12 \left[ \begin {array}{cc} -x_{1}y_{2}+{z}^{2}&-
\sqrt {2} \left( x_{1}z+x_{3} \right) \\ \noalign{\medskip}-
\sqrt {2} \left( y_{2}z+y_{3} \right) &x_{1}y_{2}-{z}^{2}\end {array} \right] .
\eea
 In these coordinates, the isomonodromic deformation equations are simply Hamiltonian's equations for the time-dependent Hamiltonians  $H_1$ and $H_2$,
modified by the ''explicit'' dependence of $L(z)$ on the deformation  parameters $(t_1, t_2)$.

Defining
\bea
x_1 &\& := u_1 {\rm e}^{w},\quad
x_2 := u_2 {\rm e}^{w},\quad 
x_3 := {\rm e}^{w}, \cr
y_1 &\& := v_1 {\rm e}^{-w},\ \ 
y_2 := v_2 {\rm e}^{-w},\ \ 
y_3 := (a - u_1 v_1 - u_2 v_2) {\rm e}^{-w},
\eea
the canonical $1$-form becomes
\bea
\theta= \sum_{i=1}^3 y_i dx_i = v_1du_1 + v_2 du_2 + a dw,
\eea
so the coordinate change from $(x_1, y_1, x_2, y_2 , x_3, y_3 )$ to $(u_1, v_1, u_2, v_2, w, a)$ is canonical.
The conserved quantity  $a$, defined in (\ref{a_H_inf}), that generates the scaling symmetry (\ref{scaling}), (\ref{scaling_Lax}) 
may also be expressed as
\be
a = \frac{1}{4}\res_{z=0} z^{-3} \tr (L^2(z)) - \frac{1}{2}t_2^2,
\ee
and  can be set equal to any constant value. The scaling flow (\ref{scaling}), (\ref{scaling_Lax})  that it generates  
 is just translation of the canonically conjugate position variable $w$ :
 \be
f_s : (u_1,v_1, u_2, v_2, w, a) \ \ra \  (u_1,v_1, u_2, v_2, w+s, a),
\ee
which is an ignorable canonical coordinate for all Hamiltonians in the ring of spectral invariants of $L(z)$:

The reduced Hamiltonians are
\bea
H_1=   &\& \left( \frac3 2v_{2}{u_{1}}^{2}-t_{2}u_{1}+2u_{2} \right) a
-2 t_{1}u_{1}v_{2}
+ \left( {u_{1}}^{2}v_{1} +u_{1}u_{2}v_{2}-v_{2} \right) t_{2} \cr
&\&-\frac 32{u_{1}}^{3}v_{1}v_{2}
-\frac 32{u_{1}}^{2}u_{2}{v_{2}}^{2}
-2u_{1}u_{2}v_{1}
+\frac 32u_{1}{v_{2}}^{2}
-2{u_{2}}^{2}v_{2}+2v_{1}
\\
H_2  = &\& \frac 1 2{a}^{2}{u_{1}}^{2}
+ \left( -u_{1}t_{1}-t_{2}-u_{1} \left( {u_{1}}^{2}v_{1}+u_{1}u_{2}v_{2}-v_{2} \right)  \right) a
+ \left( {u_{1}}^{2}v_{1}+u_{1}u_{2}v_{2}-v_{2} \right) t_{1}
+\cr
&\&+\frac  14 {t_{2}}^{2}u_{1}v_{2}
+ \left( -\frac 14{v_{2}}^{2}{u_{1}}^{2}+\frac 1 2u_{1}v_{1}+\frac 1 2u_{2}v_{2} \right) t_{2}
+\frac 12{u_{1}}^{4}{v_{1}}^{2}
+{u_{1}}^{3}u_{2}v_{1}v_{2}
+\frac 1{16}{v_{2}}^{3}{u_{1}}^{3}
+\cr
&\&+\frac 1 2{u_{1}}^{2}{u_{2}}^{2}{v_{2}}^{2}
-\frac 54{u_{1}}^{2}v_{1}v_{2}
-\frac 54 u_{1}u_{2}{v_{2}}^{2}
+\frac 1 2{v_{2}}^{2}+u_{2}v_{1}.
\eea
The isomonodromic deformation equations are then Hamiltonian's equations for the time-dependent Hamiltonians  $H_1$ and $H_2$.

%%%%%%%%%%%%%%%%%%%%% Acknowledgements %%%%%%%%%%%%%%%%%
\bigskip
\bigskip
\noindent 
{\small{\it Acknowledgements.} We would like thank G\'abor Pusztai, who contributed valuably to the earlier stage 
of this project (in 2003-2004), and Philip Boalch, for bringing our attention to  ref.~\cite{Yam}.
 This work  was supported by the Natural Sciences and Engineering Research Council of Canada (NSERC). 
\bigskip

 %%%%%%%%%%%%%%%%%%%%% Data availability %%%%%%%%%%%%%%%%%
\smallskip
\noindent
 \small{\it Data availability.} Data sharing is not applicable to this article since no data are involved. 
 
   %%%%%%%%%%%%%%%%%%%%% Bibliography %%%%%%%%%%%%%%%%% 

\end{document}